\documentclass[12pt]{article}
\usepackage{ametsoc}
\usepackage{graphicx}
\usepackage{natbib}

\newcommand{\myabstract}{The effect of Prandtl number on the evolution of Kelvin-Helmholtz (KH) billows and the amount of mixing they generate is studied through direct numerical simulation (DNS). The results indicate that the time evolution of the rate of mixing  through different stages of the life-cycle of KH flow is significantly influenced by the Prandtl number. As the Prandtl number increases, the final amount of mixing increases for Reynolds that are too low to support active three-dimensional motions. This trend is the opposite in sufficiently high Reynolds number KH flows that can overcome viscous effects, and develop significant three-dimensional instabilities. While the mixing generated in the two-dimensional flows, uniform in the span-wise direction, is not significantly dependent on the Prandtl number, the turbulent mixing induced by three-dimensional motions is a function of the Prandtl number. The turbulent mixing efficiency near the end of the turbulence decay phase approaches 0.2, the commonly observed value in the ocean. A smooth transition in mixing is observed by increasing the buoyancy Reynolds number. } 
\begin{document}
%
%%%%%%%%%%%%%%%%%%%%%%%%%%%%%%%%%%%%%%%%%%%%%%%%%%%%%%%%%%%%%%%%%%%%%
% TITLE
%
% Enter your TITLE here
%%%%%%%%%%%%%%%%%%%%%%%%%%%%%%%%%%%%%%%%%%%%%%%%%%%%%%%%%%%%%%%%%%%%%
\title{\textbf{\large{The effect of Prandtl number on mixing in low Reynolds number Kelvin-Helmholtz billows}}}
%
% Author names, with corresponding author information. 
% [Update and move the \thanks{...} block as appropriate.]
%
\author{\textsc{Mona Rahmani $^{1}$,}
				\thanks{\textit{Corresponding author address:} 
				Mona Rahmani, Department of Civil Engineering,
				6250 Applied Science Lane, Vancouver, BC, Canada, V6T 1Z4. 
				\newline{E-mail: mona.rahmani80@gmail.com}}\quad\textsc{Gregory Lawrence $^{1}$, and Brian Seymour $^{2}$}\\
\textit{\footnotesize{$^{1}$ Department of Civil Engineering, University of British Columbia, Vancouver, BC, Canada}}\\
\textit{\footnotesize{$^{2}$ Department of Mathematics, University of British Columbia, Vancouver, BC, Canada}}
\and 
%\centerline{\textsc{Extra Author}}\\% Add additional authors, different insitution
%\centerline{\textit{\footnotesize{Affiliation, City, State/Province, Country}}}
}
%
% The following block of code will handle the formatting of the title page depnding on whether
% we are formatting a double column (dc) author draft or a single column paper for submission.
% AUTHORS SHOULD SKIP OVER THIS... There is nothing to do in this section of code.
\ifthenelse{\boolean{dc}}
{
\twocolumn[
\begin{@twocolumnfalse}
\amstitle

% Start Abstract (Enter your Abstract above.  Do not enter any text here)
\begin{center}
\begin{minipage}{13.0cm}
\begin{abstract}
	\myabstract
	\newline
	\begin{center}
		\rule{38mm}{0.2mm}
	\end{center}
\end{abstract}
\end{minipage}
\end{center}
\end{@twocolumnfalse}
]
}
{
\amstitle
\begin{abstract}
\myabstract
\end{abstract}
\newpage
}
%%%%%%%%%%%%%%%%%%%%%%%%%%%%%%%%%%%%%%%%%%%%%%%%%%%%%%%%%%%%%%%%%%%%%
% MAIN BODY OF PAPER
%%%%%%%%%%%%%%%%%%%%%%%%%%%%%%%%%%%%%%%%%%%%%%%%%%%%%%%%%%%%%%%%%%%%%
%

\section{Introduction}
Turbulent breakdown of Kelvin-Helmholtz (KH) instabilities is an important source of mixing in the ocean and atmosphere.  In these environments, the Prandtl number, the ratio of viscosity to molecular diffusivity of scalars, varies over a wide range from nearly 0.7 for heat in the atmosphere up to 700 for salt in the ocean. However, the effect of Prandtl number on mixing is still not well understood. This is mainly because high Prandtl number mixing occurs at very small scales and requires high-resolution numerical simulations or field and laboratory measurements. While our ability to model the effect of high Prandtl number on mixing at oceanic and atmospheric Reynolds numbers is still out of our reach, examining the Prandtl number effects at low and moderate Reynolds numbers may still provide useful information with respect to oceanic and atmospheric processes.

The amount of entrainment, and therefore mixing, in KH billows is controlled by large and small-scale vortical structures, as demonstrated in the experiments of \cite{brownetal1974,breidenthal1981,thorpe1985,bernaletal1986,pattersonetal2006}, and field measurements of \cite{seimetal1994,geyeretal2010}. In addition to the primary two-dimensional KH vortex, an important mechanism for entrainment is the growth of three-dimensional span-wise instabilities. These three-dimensional motions in KH billows are sustained above sufficiently high Reynolds numbers, creating a "mixing transition" \cite[]{konrad1976,breidenthal1981,koochesfahanietal1986,dimotakis2005,rahmani2011c}. Experiments by  \cite{konrad1976} at Prandtl number of 0.7 and  \cite{breidenthal1981} and \cite{koochesfahanietal1986} at Prandtl number of 600 exhibited the same range of transitional Reynolds numbers, but the experiments at the higher Prandtl number exhibited significantly less mixing.

%The transition in mixing starts at outer-scale Reynolds numbers of $O(10^{3})$, and completes by reaching a Reynolds number independent regime at outer-scale Reynolds numbers of $O(10^{4})$. When stated in terms of the initial Reynolds numbers, this transition range is between $O(10^2)$ and $O(10^{3})$ \cite[]{breidenthal1981}.

%Comparison of the mixing transition observed by \cite{konrad1976} at Prandtl number of 0.72 with that observed by \cite{breidenthal1981} and \cite{koochesfahanietal1986} at Prandtl number of 600 suggests that increasing the Prandtl number does not affect the range of the transitional Reynolds numbers, but does reduce the amount of mixing.  Therefore, when studying Prandtl number effects it is important to know the regime of the flow with respect to the mixing transition.

Subsequent studies have shown that the relationship between mixing and Prandtl number is governed by the competition between enhanced interfacial area and slower rate of diffusion at higher Prandtl numbers \cite[]{cortesietal1999,staquet2000}.  However, interpretation of direct effects of Prandtl number on mixing is potentially complicated by other factors including: double-diffusion, vortex pairing, varying Richardson number, or the limited range of Prandtl numbers studied. The flow exhibits finer structures at high Prandtl numbers \cite[]{rahmani2011c,balmforth2012,rahmani2014}, which potentially influence the transition to turbulence. The effect of Prandtl number on enhancing three-dimensionality was indicted in the studies of \cite{klaassenetal1985b,cortesietal1998,cortesietal1999,staquet2000,mashayek2012}. Direct numerical simulation (DNS) of KH flows by \cite{cortesietal1998,cortesietal1999} show that by increasing the Prandtl number from 0.00535 to 2.2, three-dimensional motions and interfacial area are enhanced, and therefore higher entrainment and mixing rates are obtained. However, in DNS of \cite{staquet2000}, when Prandtl number increased from 0.7 to 1.4, the rate of mixing increased or decreased depending on the Richardson number. In double-diffusive scalars with Prandtl numbers of 7 and 50, using DNS, \cite{smythetal2005} showed that the rate of turbulent mixing is lower for the scalar with higher Prandtl number. 

\begin{figure}
\centering
\includegraphics[width=12.5cm]{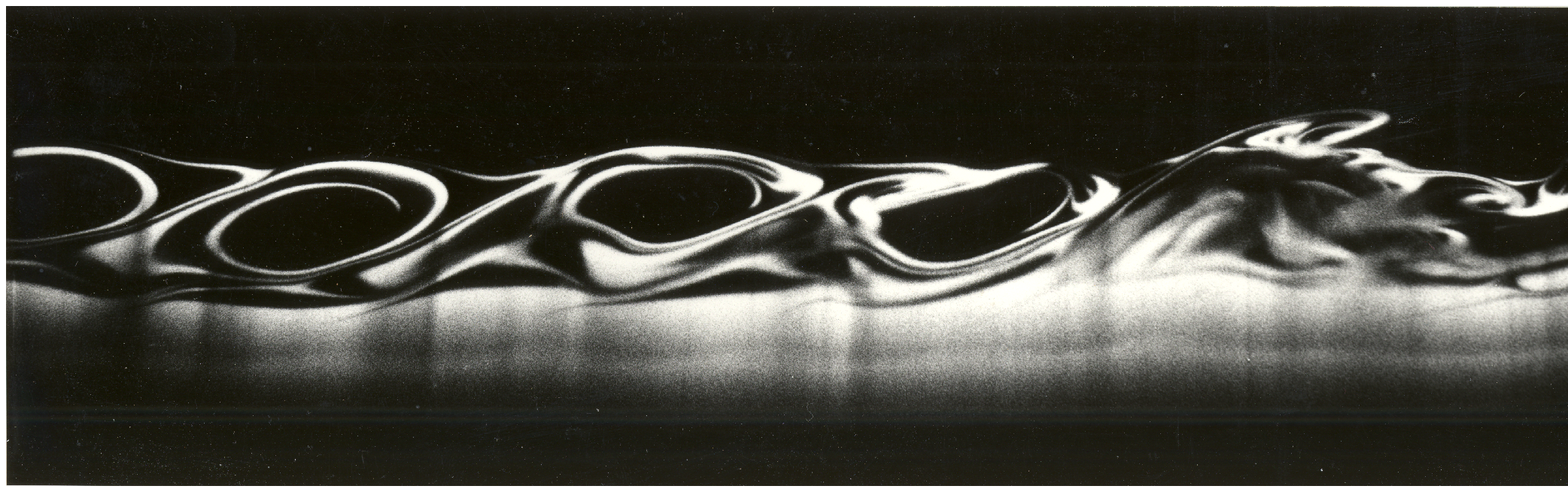}
\includegraphics[width=12.5cm]{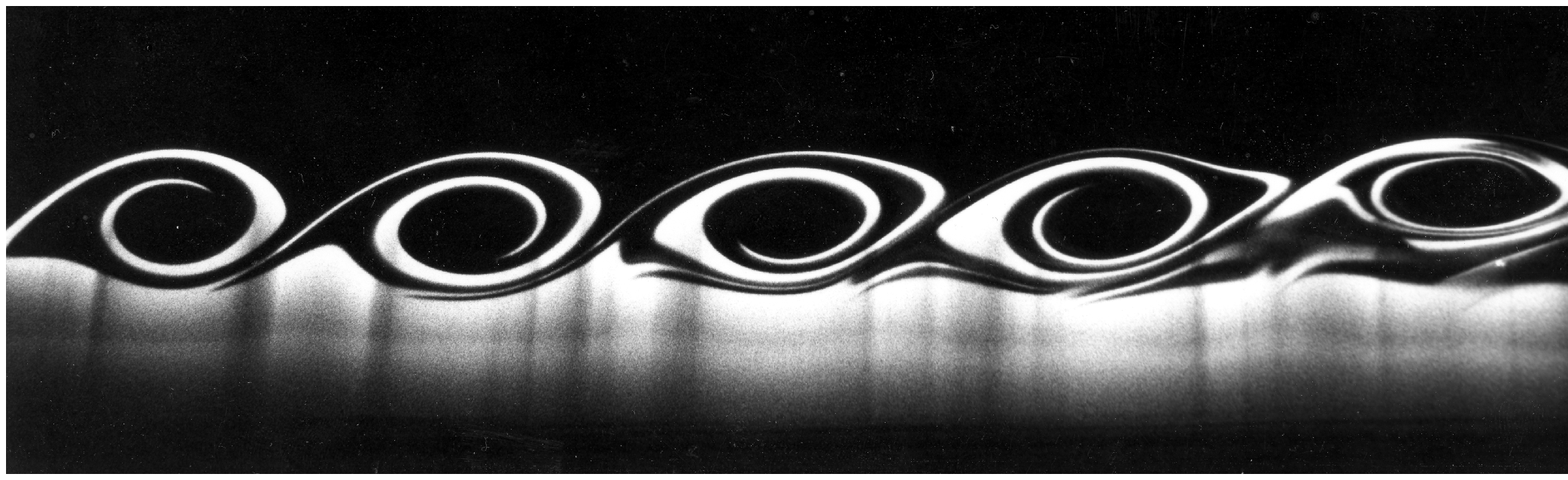}
\caption{Evolution and breakdown of KH billows in laboratory experiments of Lawrence.}
\label{fig:exp}
\end{figure}

In the present work, our objective is to use DNS to examine the competing effects of increased interfacial area, but a reduced rate of molecular mixing as Prandtl number increases on both two and three-dimensional mixing throughout the lifetime of KH billows. We investigate a range of Prandtl numbers from 1-700 for Reynolds numbers below and within the mixing transition. We restrict our attention to individual billows, as they are the dominant flow feature in the laboratory experiments of KH billows generated by splitter plates (see figure \ref{fig:exp}), and a single Richardson number; the effects of vortex pairing of KH billows  \cite[e.g. see][]{moseretal1991,winantetal1974} and differential strengths of stratification \cite[e.g. see][]{cortesietal1999,caulfieldetal2000} are the subject of future work.

\section{Numerical simulation}
The Boussinesq Navier-Stokes equations are solved using the model described by \cite{wintersetal2004} and \cite{smythetal2005}. We choose an initial flow with horizontal velocity and density profiles defined by
\begin{equation}
U(z)=\frac{\Delta U}{2} \textnormal{tanh}(\frac{2z}{\delta_{0}}), \; \; \; \; \;  \textnormal{ and }  \; \; \; \; \;
%\label{eq4}
%\end{equation}
%\begin{equation}
\bar{\rho}(z)=-\frac{\Delta \rho}{2} \textnormal{tanh}(\frac{2z}{h_{0}}),
\label{eqn:profiles}
\end{equation}
\noindent with $\delta_{0}$ being the initial value for the time-varying vorticity interface thickness of $\delta$, $\textit{h}_{0}$ the initial value for the time-varying density interface thickness of $h$, $\Delta U$ the horizontal velocity difference and $\Delta \rho$ the density difference. The dimensionless numbers for this flow are
\begin{equation}
Re_{0}=\frac{\delta_{0}\Delta U}{\nu}, \; \; \; \textit{J}=\frac{\Delta \rho g \delta_{0}}{\rho _0 \Delta U^{2}}, \; \; \; Pr=\frac{\nu}{\kappa}, \; \; \; \textnormal{ and }\; R=\frac{\delta_{0}}{h_{0}}, \; \; \;\
\label{eqn:nond_param}
\end{equation}
\noindent the initial Reynolds number, bulk Richardson number, Prandtl number and scale ratio, where $\nu$ denotes the kinematic viscosity, $\kappa$ the molecular diffusivity, $\rho_{0}$ the reference density and $g$ the gravitational acceleration. The scale ratio is related to the Prandtl number by $R=Pr^{1/2}$. The general flow field in a cartesian coordinate system of $(x,y,z)$ has a velocity field of $(u,v,w)$ and density field of $\rho$, where
\begin{equation}
u=U(z)+u^{'}(x,y,z), \; \; \; v=v^{'}(x,y,z), \; \; \; w=w^{'}(x,y,z), \; \; \; \rho=\bar{\rho}(z)+\rho^{'}(x,y,z), \; \; \;\
\label{eqn:fields}
\end{equation}

\noindent with $(u^{'},v^{'},w^{'})$ and $\rho^{'}$ being the perturbation velocity and density fields.

The choice of simulation parameters, shown in table \ref{tab:sim}, is as follows. Different Prandtl number simulations ($Pr$ = 1, 9, 16, 25, and 64) are performed in three dimensions at $Re_{0}$ = 100, 300, 400, and 600.  Simulation of $Pr$ = 64 and $Re_{0}$ = 600 flow was not possible due to numerical resources limitations. Simulations at $Re_{0}$ = 300, 400, and 600 are also performed in two dimensions to distinguish between the role of three-dimensional motions versus two-dimensional flow features in the generation of mixing. At $Re_{0}$ = 300, a two-dimensional simulation was also performed at $Pr$ = 700, representing the diffusion of salt water in the ocean. In all simulations, one wavelength of the fastest growing mode of the instability, $L_{x}$, is considered.  The height of the domain is $L_{z}=9\delta_{0}$, and the width of the domain in three-dimensional simulations is $L_{y}=L_{x}/2$, large enough to accommodate at least one wavelength of span-wise instabilities. Periodic horizontal and free-slip vertical boundary conditions are applied. The basic velocity and density profiles given in equations \ref{eqn:profiles} are perturbed using the eigenfunctions from the numerical solution of the linear stability equation. The amplitude of the two-dimensional perturbation had a negligible effect on mixing. Also, a random noise is added to the velocity field to initiate three-dimensional secondary instabilities. In all simulations the grid spacing is smaller than 2.6$L_{B}$, where $L_{B}$ is the Batchelor length scale \cite[]{batchelor1959}, a length scale below which density gradients are suppressed by molecular diffusion.

\section{Two and three-dimensional evolution of KH billows}
The evolution of  single KH billows in laboratory experiments of Lawrence are shown in figure \ref{fig:exp}.  In these experiments, a faster flowing fresh water top layer flows on a slower flowing salt water. After the roll-up, and evolution in two dimensions, the billows become chaotic through the development of small-scale motions. This figure illustrates the sharp interfaces between the two layers at the high Prandtl number.  

\begin{figure}[t]
%\centering
%\hspace{0.2cm}
\begin{minipage}[h]{0.23\linewidth} % A minipage that covers half the page
%\centering
\includegraphics[height=3.8cm]{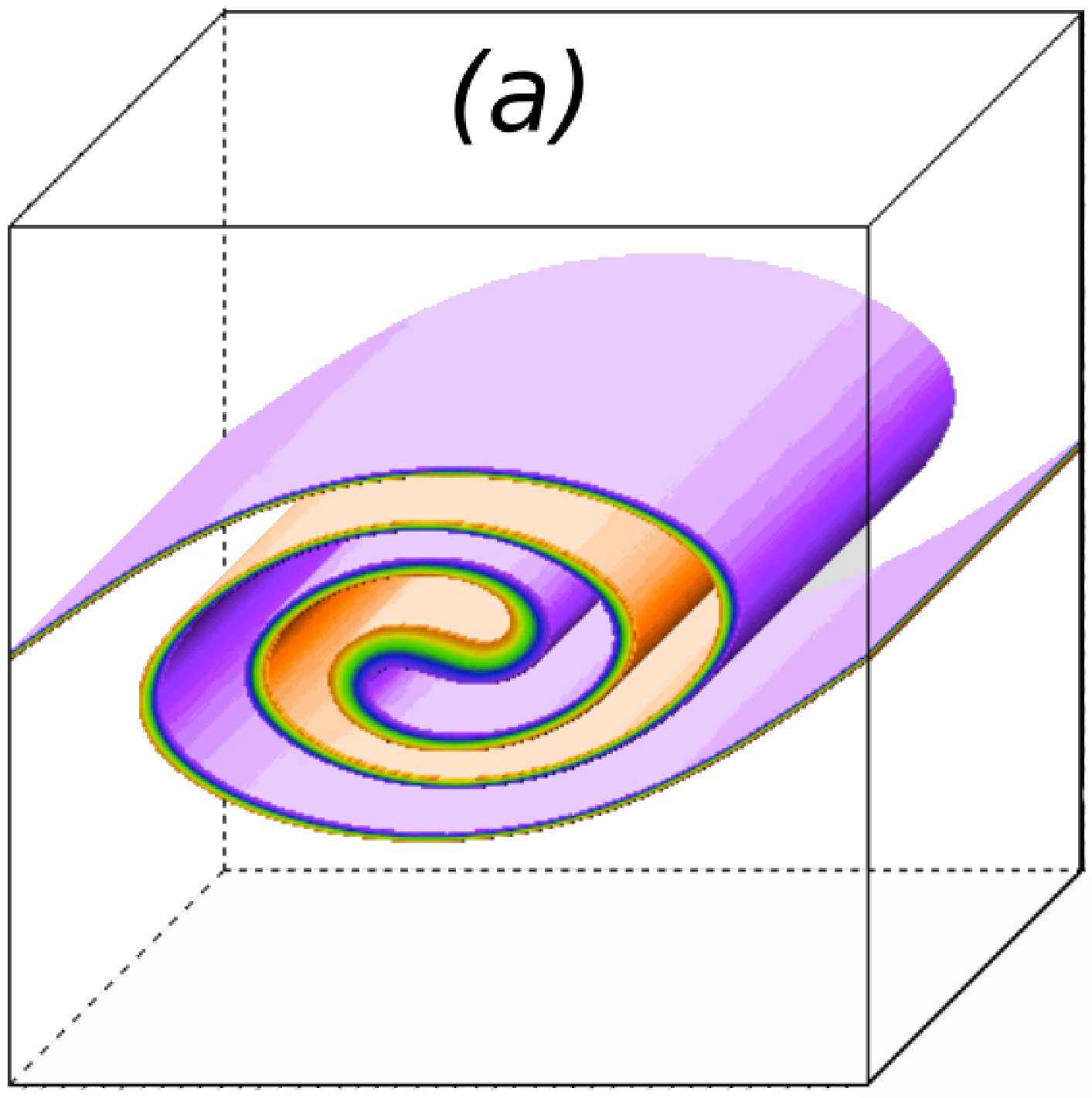}
\end{minipage}
%\centering
\hspace{0.1cm} % To get a little bit of space between the figures
\begin{minipage}[h]{0.23\linewidth} % A minipage that covers half the page
%\centering
\includegraphics[height=3.8cm]{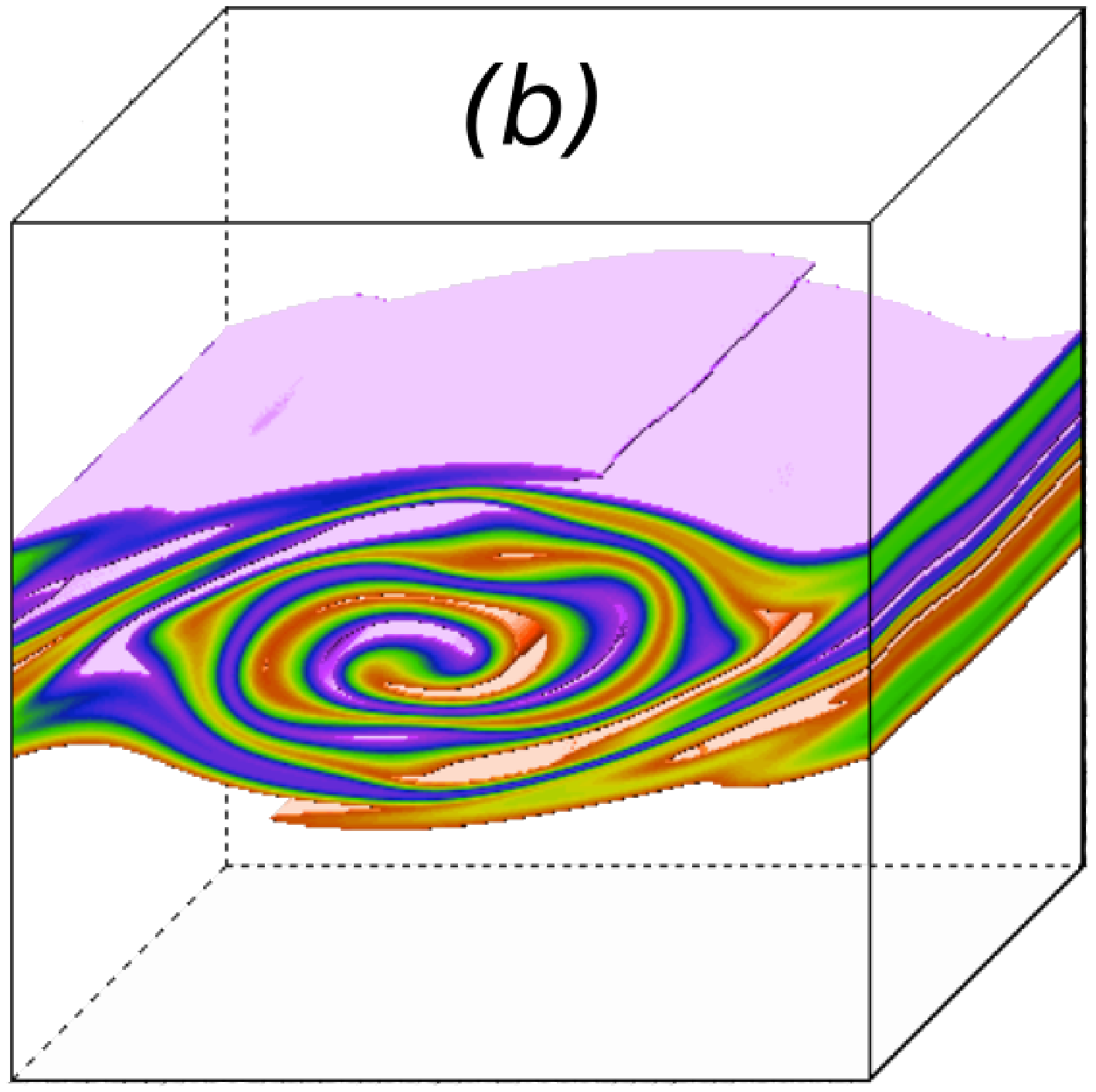}
\end{minipage}
\hspace{0.1cm} % To get a little bit of space between the figures
\begin{minipage}[h]{0.23\linewidth} % A minipage that covers half the page
%\centering
\includegraphics[height=3.8cm]{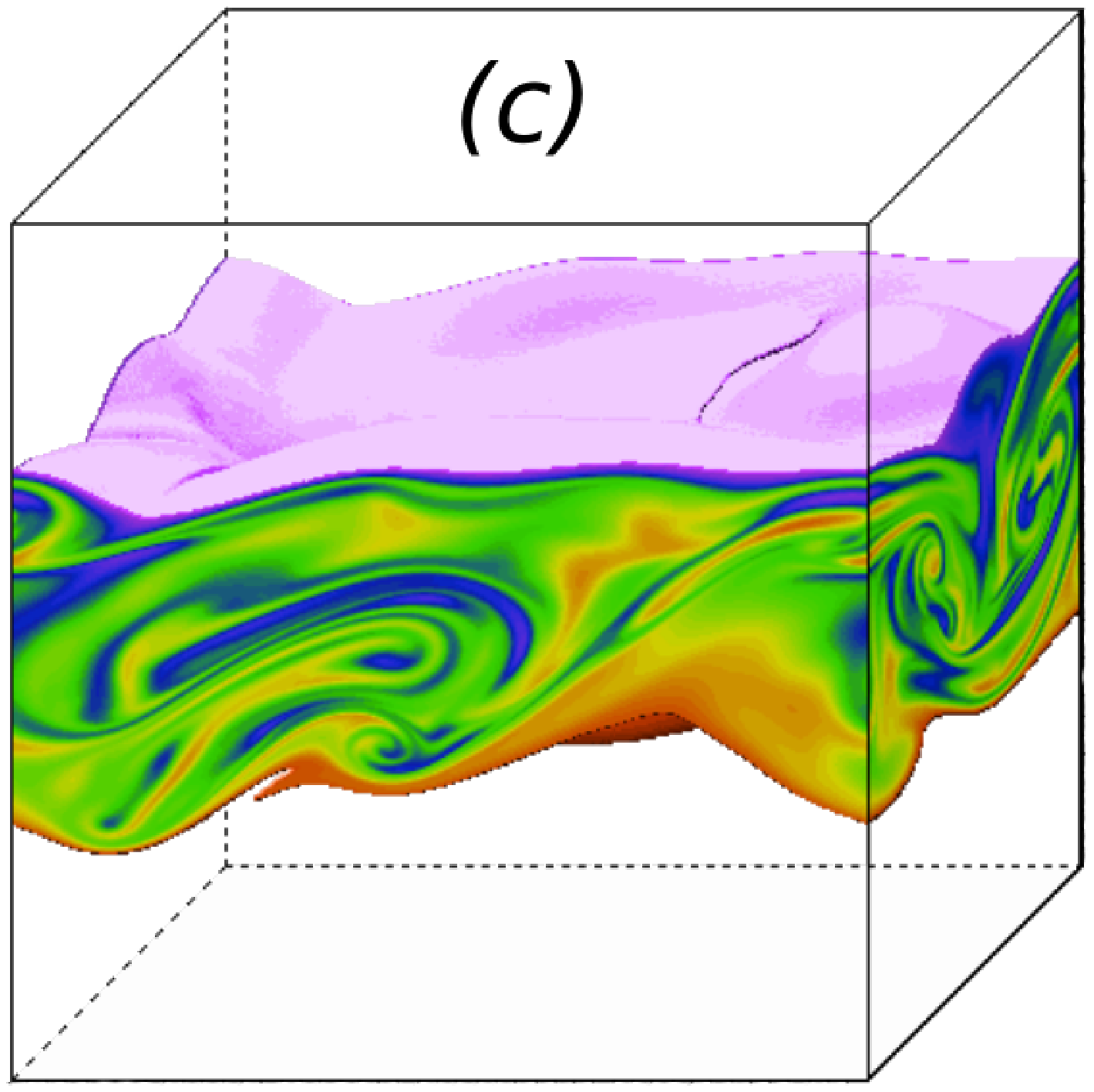}
\end{minipage}
\hspace{0.1cm} % To get a little bit of space between the figures
\begin{minipage}[h]{0.23\linewidth} % A minipage that covers half the page
%\centering
\includegraphics[height=3.8cm]{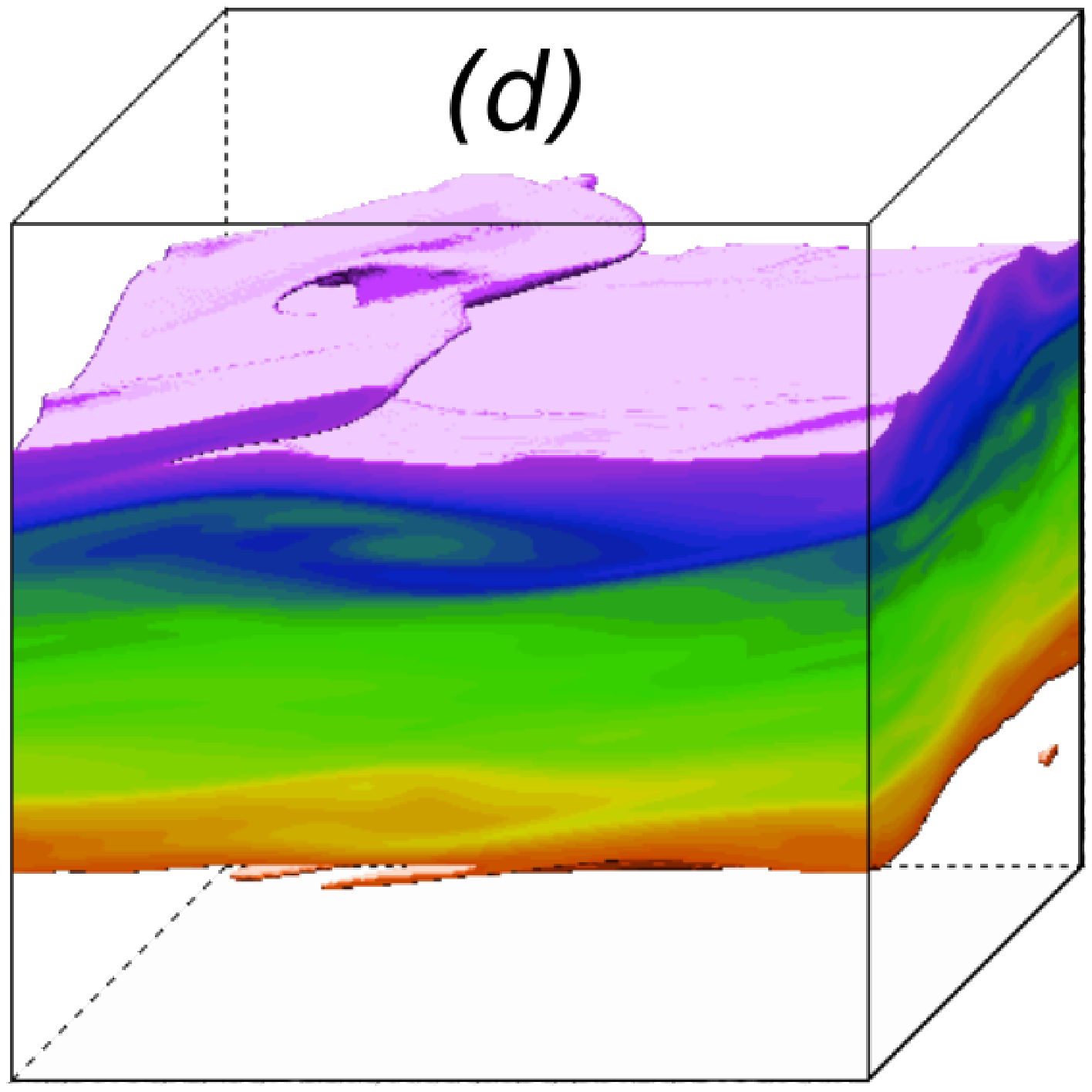}
\end{minipage}
\begin{minipage}[h]{0.95\linewidth} % A minipage that covers half the page
\centering
\includegraphics[height=5.8  cm]{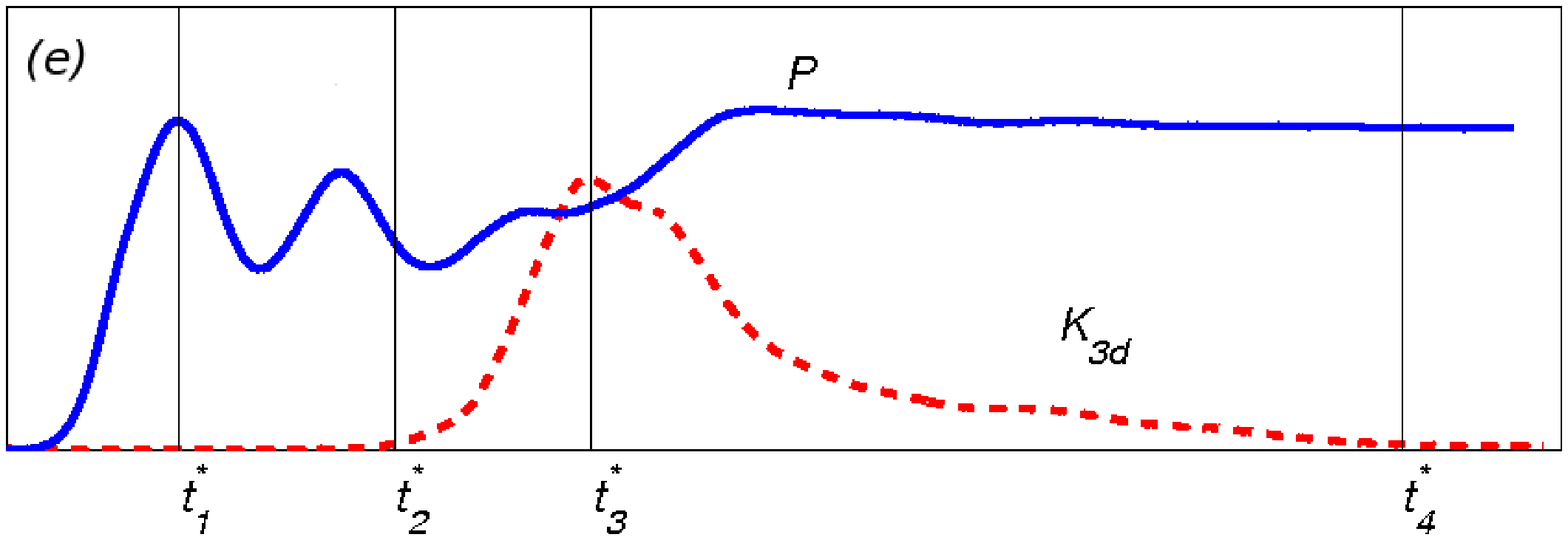}
\centering
\end{minipage}
\caption[The density structure of a KH billow at different stages]{The density structure of a KH billow at the end of each phase, $t^{*}_{1}$, $t^{*}_{2}$, $t^{*}_{3}$ and $t^{*}_{4}$, in panels (a), (b), (c) and (d), respectively, for $Re_{0}$ = 400 and $Pr$ = 25. These times correspond to the maximum of the potential energy, the onset of three-dimensional instability, the maximum intensity of three-dimensional motions and re-laminarization, respectively. The definition of the different stages is illustrated in panel (e) using the total potential energy, $P$, and the three-dimensional turbulent kinetic energy, $K_{3d}$.}
\label{fig:dens_evolution}
\end{figure}

Following \cite{caulfieldetal2000}, we divide the life-cycle of a KH instability into four phases. The end of each phase is marked by a transition time, $t^{*}_{1}$, $t^{*}_{2}$, $t^{*}_{3}$, or $t^{*}_{4}$, with $t^{*}$ = $t\Delta U /\delta_{0}$ being the non-dimensional time. The three-dimensional density structure of a simulation with $Re_{0}$ = 400 and $Pr$ = 25 at these four transitions is shown in figure \ref{fig:dens_evolution}. The first phase is the two-dimensional growth of the billow, ending at $t^{*}_{1}$, when a maximum billow height and therefore total potential energy, $P$, is reached. The second phase is the subsequent two-dimensional evolution of the billow, ending at $t^{*}_{2}$ when secondary three-dimensional instabilities start to grow. We quantitatively mark  $t^{*}_{2}$ as the time when $K_{3d}/K_{0}>10^{-3}$ for the first time, where $K_{3d}$ is the kinetic energy of the three-dimensional motions, and $K_{0}$ the initial kinetic energy of the flow.  Up to $t^{*}_{2}$ two and three-dimensional simulations have identical flow fields. The third phase is the growth of three-dimensional motions until reaching a maximum intensity, measured by $K_{3d}$ reaching a maximum at $t^{*}_{3}$. The fourth phase is the decay of three-dimensional motions and destruction of the billow until a laminar flow is reached at $t^{*}_{4}$, the time when $K_{3d}/K_{0}$ drops below $10^{-3}$.

\begin{figure}[t]
%\vspace{-1.5cm}
%\hspace{5cm}
\hspace{-1.0cm} % To get a little bit of space between the figures
\begin{minipage}[h]{0.15\linewidth} % A minipage that covers half the page
\centering
\includegraphics[height=2.85cm]{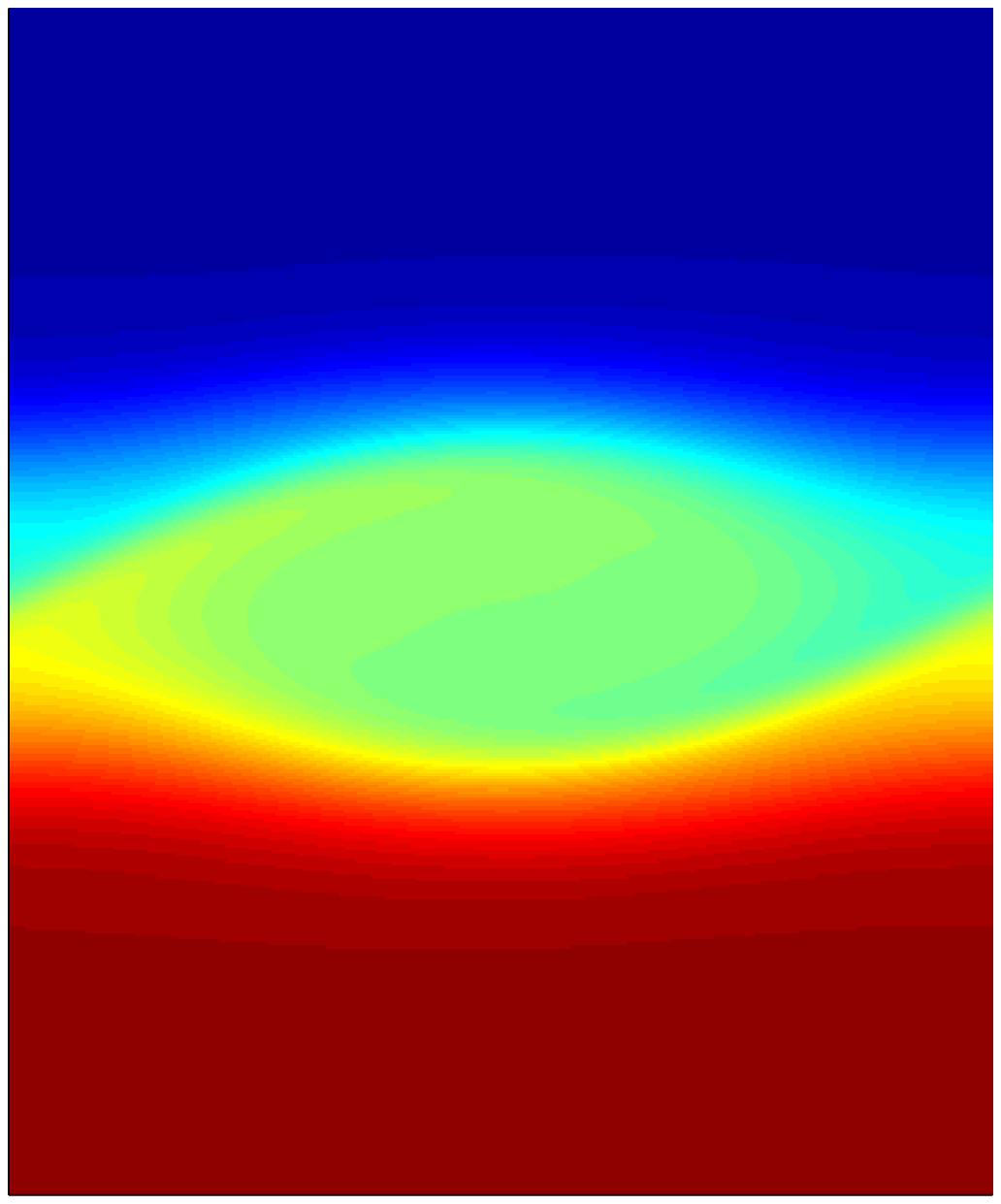}
\end{minipage}
\centering
\hspace{-0.3cm} % To get a little bit of space between the figures
\begin{minipage}[h]{0.15\linewidth} % A minipage that covers half the page
\centering
\includegraphics[height=2.85cm]{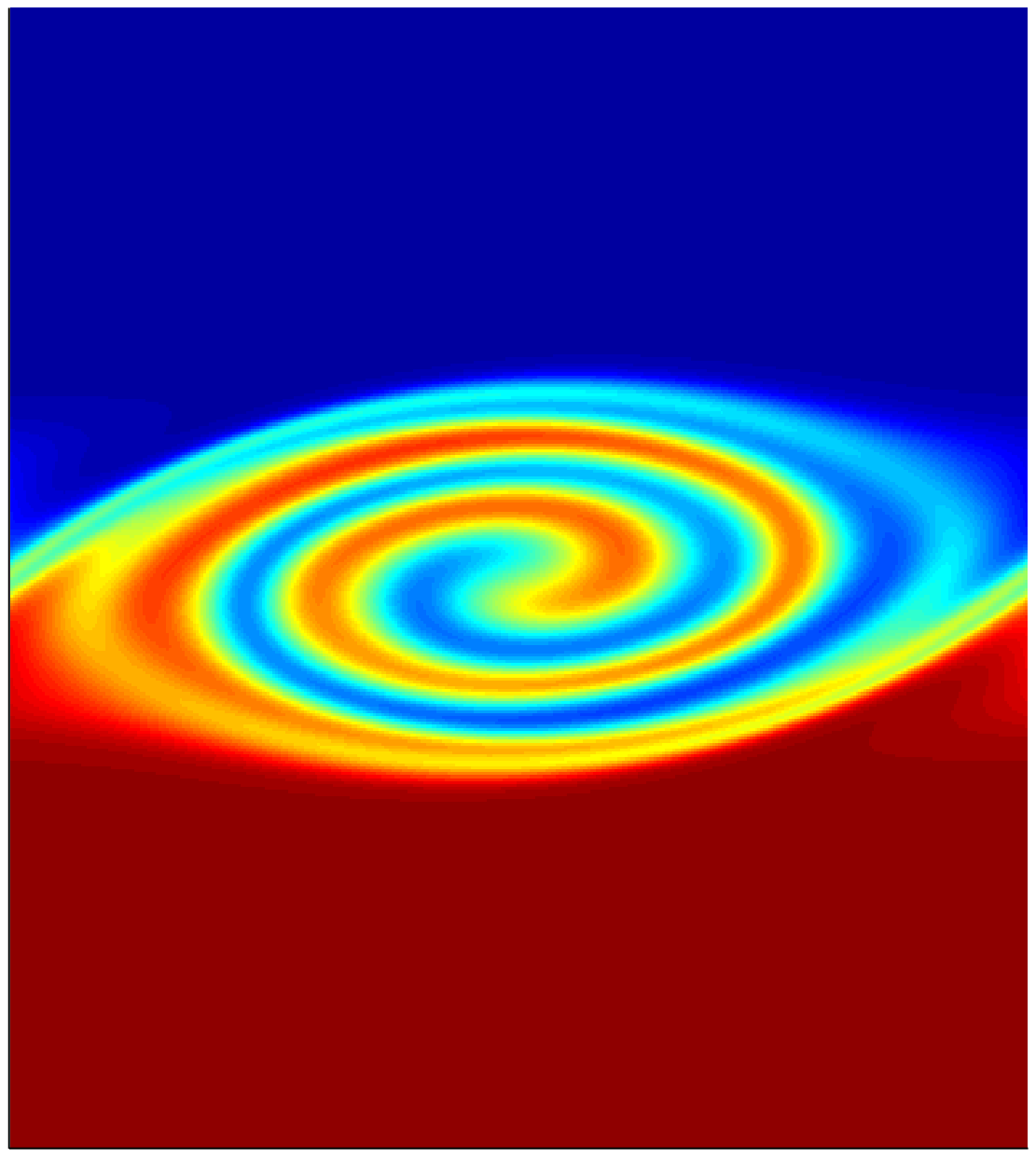}
\end{minipage}
\centering
\hspace{-0.05cm} % To get a little bit of space between the figures
\begin{minipage}[h]{0.15\linewidth} % A minipage that covers half the page
\centering
\includegraphics[height=2.85cm]{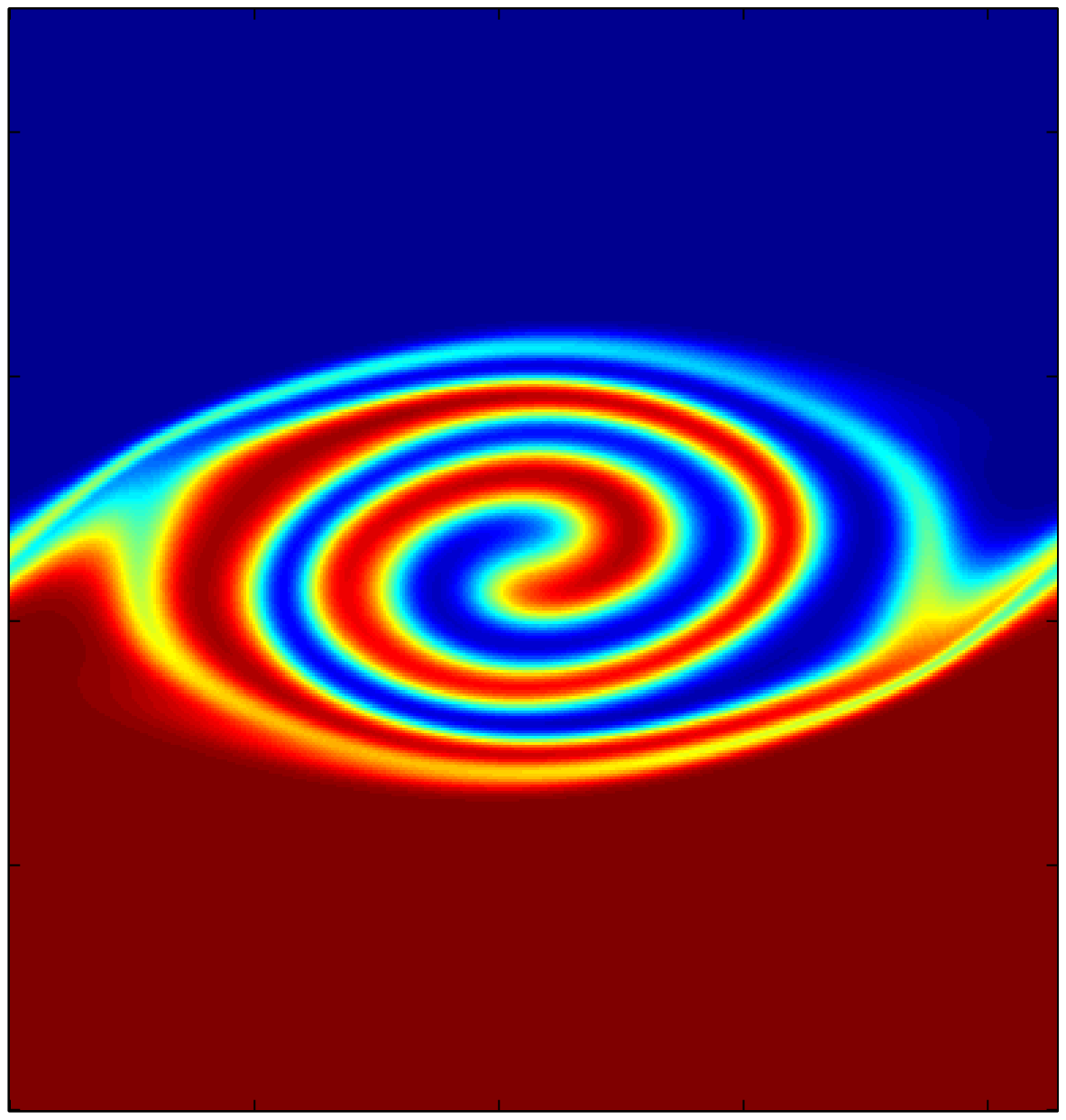}
\end{minipage}
\centering
\hspace{0.05cm} % To get a little bit of space between the figures
\begin{minipage}[h]{0.15\linewidth} % A minipage that covers half the page
\centering
\includegraphics[height=2.85cm]{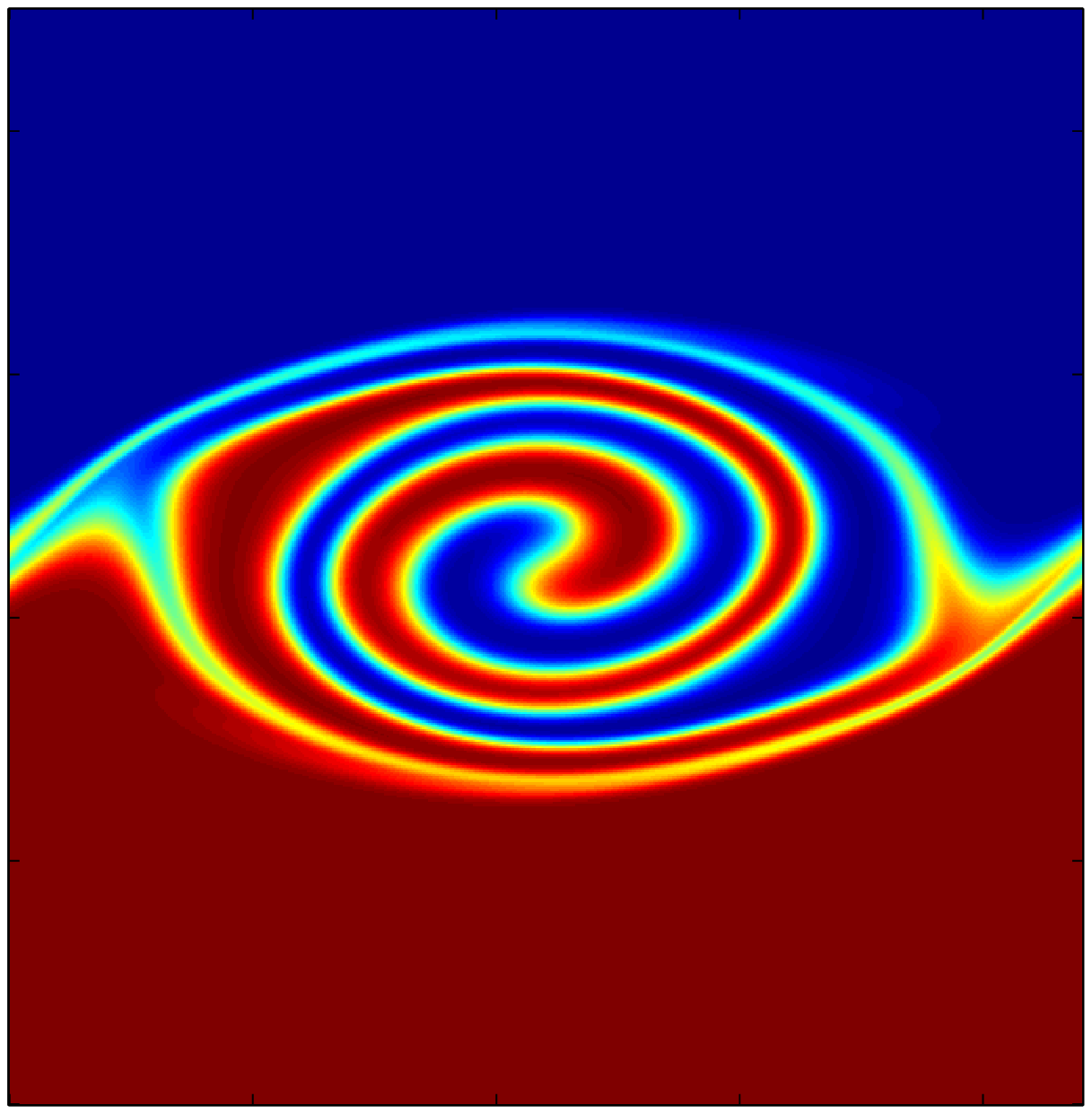}
\end{minipage}
\centering
\hspace{0.05cm} % To get a little bit of space between the figures
\begin{minipage}[h]{0.15\linewidth} % A minipage that covers half the page
\centering
\includegraphics[height=2.85cm]{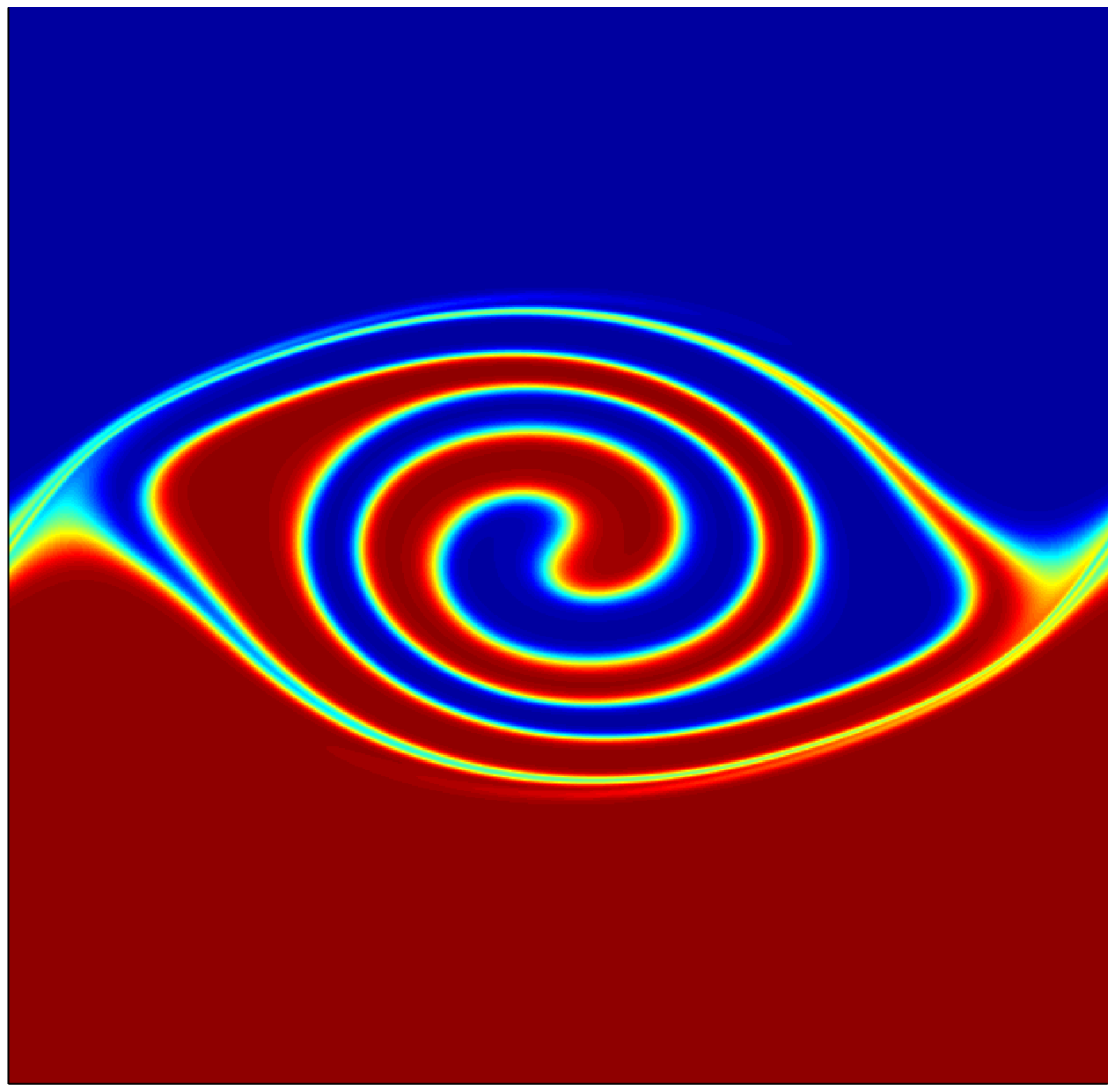}
\end{minipage}
\centering
\hspace{0.15cm} % To get a little bit of space between the figures
\begin{minipage}[h]{0.15\linewidth} % A minipage that covers half the page
\centering\includegraphics[height=2.85cm]{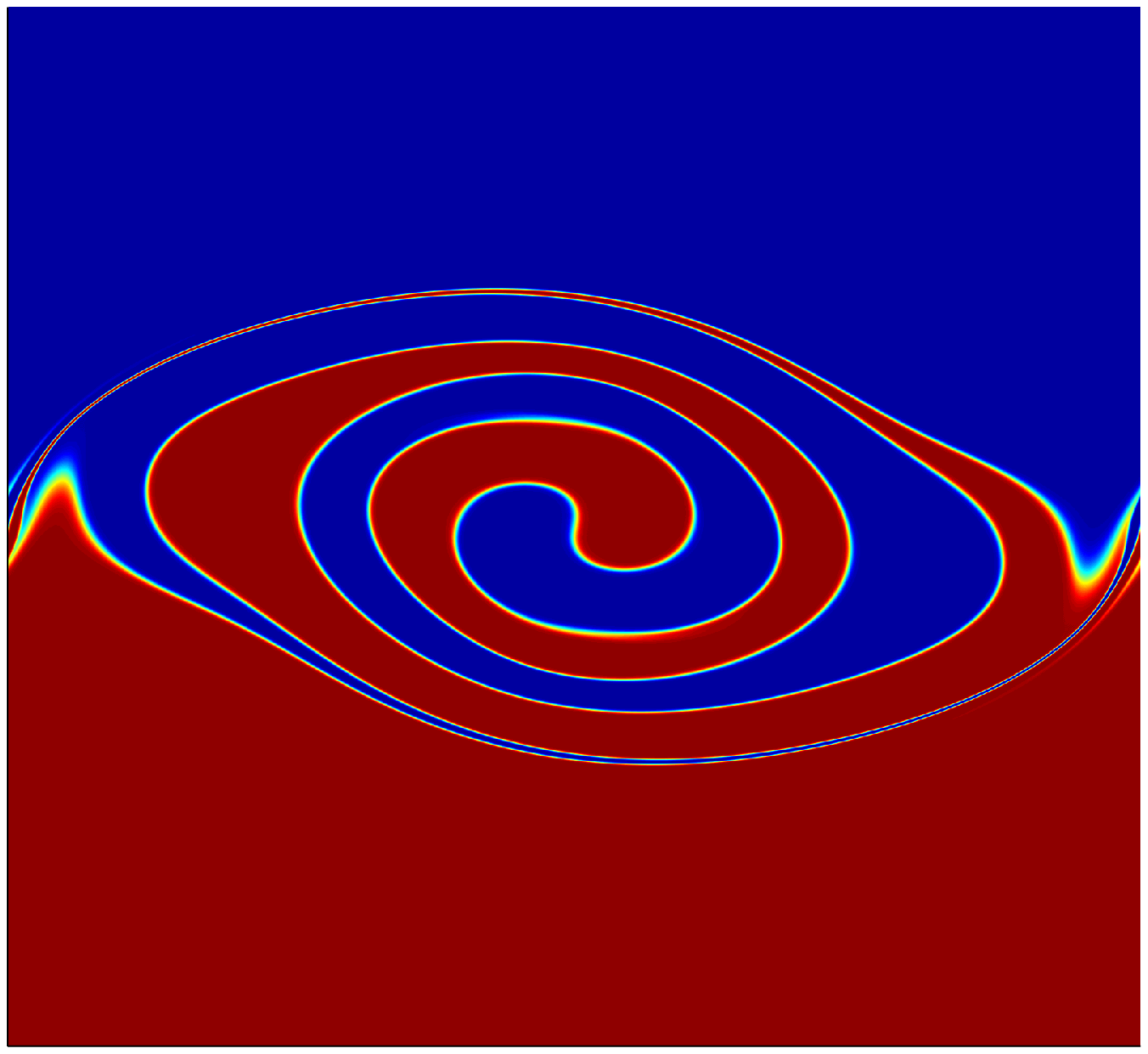}
\end{minipage}
%\centering
%\hspace{-0.4cm} % To get a little bit of space between the figures
%\begin{minipage}[h]{0.12\linewidth} % A minipage that covers half the page
%\centering
%\includegraphics[height=3.0cm]{./snapshots_old/run1_stream_3D_t_3.eps}
%\end{minipage}
%\centering
%\hspace{-0.69cm} % To get a little bit of space between the figures
%\begin{minipage}[h]{0.12\linewidth} % A minipage that covers half the page
%\centering
%\includegraphics[height=3.0cm]{./snapshots_old/run3_stream_3D_t_3.eps}
%\end{minipage}
%\centering
%\hspace{-0.4cm} % To get a little bit of space between the figures
%\begin{minipage}[h]{0.12\linewidth} % A minipage that covers half the page
%\centering
%\includegraphics[height=3.0cm]{./snapshots_old/run4_stream_3D_t_3.eps}
%\end{minipage}
%\centering
%\hspace{-0.08cm}
%\begin{minipage}[h]{0.12\linewidth} % A minipage that covers half the page
%\centering\includegraphics[height=3.0cm]{./snapshots_old/run5_t_170_2d_new_v06.eps}
%\end{minipage}
\caption[Snapshots of the stream-wise density structure of the billow]{Snapshots of the stream-wise density structure of the billow at $y=L_{y}/2$, at time $t^{*}_{2}$ at $Re_{0}$ = 300 for $Pr$ = 1, 9, 16, 25, 64 and 700 from left to right.}
\label{snapshot_stream}
\end{figure}

Snapshots of the stream-wise structure of billowing of three-dimensional KH instabilities for different Prandtl numbers at $Re_{0}$ = 300, at $t^{*}_{2}$ are shown in figure \ref{snapshot_stream}.  As $Pr$ increases, the billows exhibit more structured small-scale features that are absent in the density structure of low $Pr$ billows due to the rapid diffusion. At $Pr$ = 1, the density structure of the billow has been mainly destroyed by diffusion at $t^{*}_{2}$. The highly structured density field of the $Pr=700$ billow resembles the density field of the billows observed in the experiments of \cite{schowalteretal1994} with $Pr\sim$ 1500 and \cite{atsavapraneeetal1997} with $Pr\sim$ 600, where the billow consists of layers of rolling fluid with very sharp interfaces.  At $Pr=700$, very fine structures have formed, with very little diffusion. These structures form layers of diffusing fluid that enhance the available interfacial area between the layers, and therefore mixing.

\begin{figure}[t]
%\vspace{-1.5cm}
\hspace{-0.5cm}
\centering
\begin{minipage}[h]{0.18\linewidth} % A minipage that covers half the page
\centering
\includegraphics[height=6.0cm]{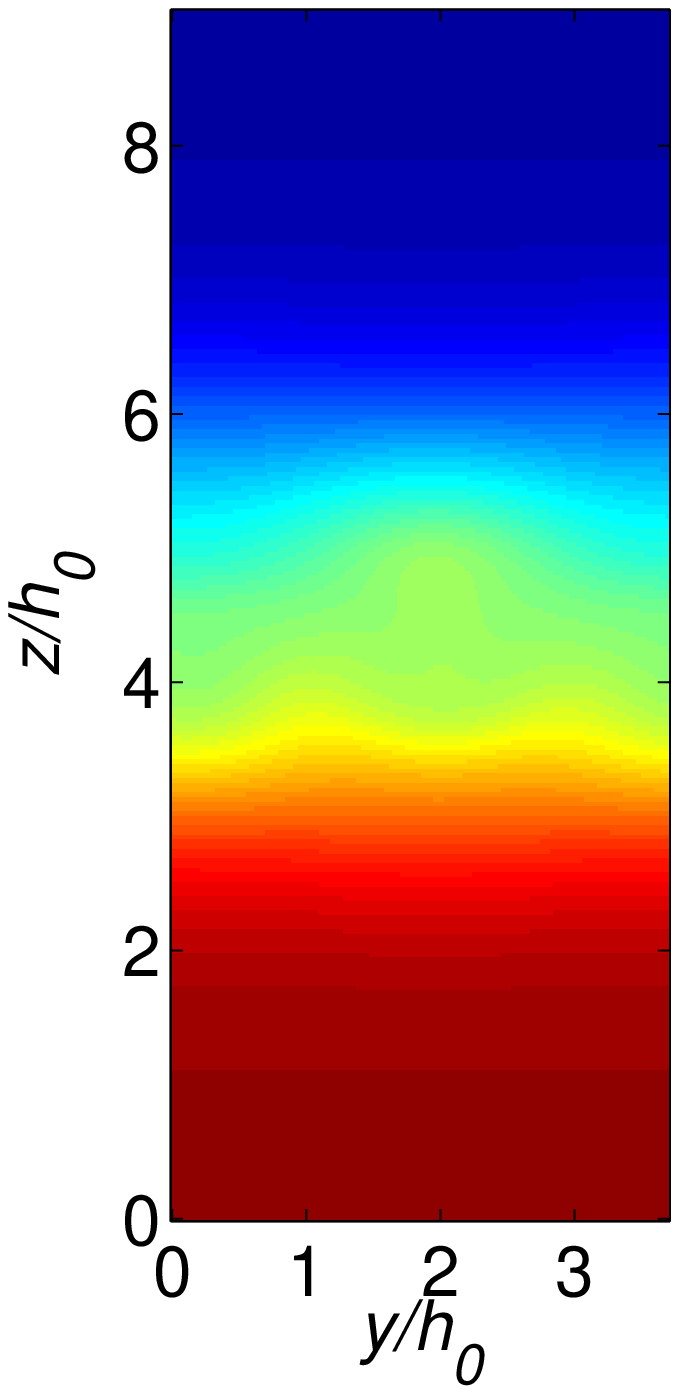}
\end{minipage}
\centering
\hspace{0.3cm} %
\begin{minipage}[h]{0.18\linewidth} % A minipage that covers half the page
\centering
\includegraphics[height=6.0cm]{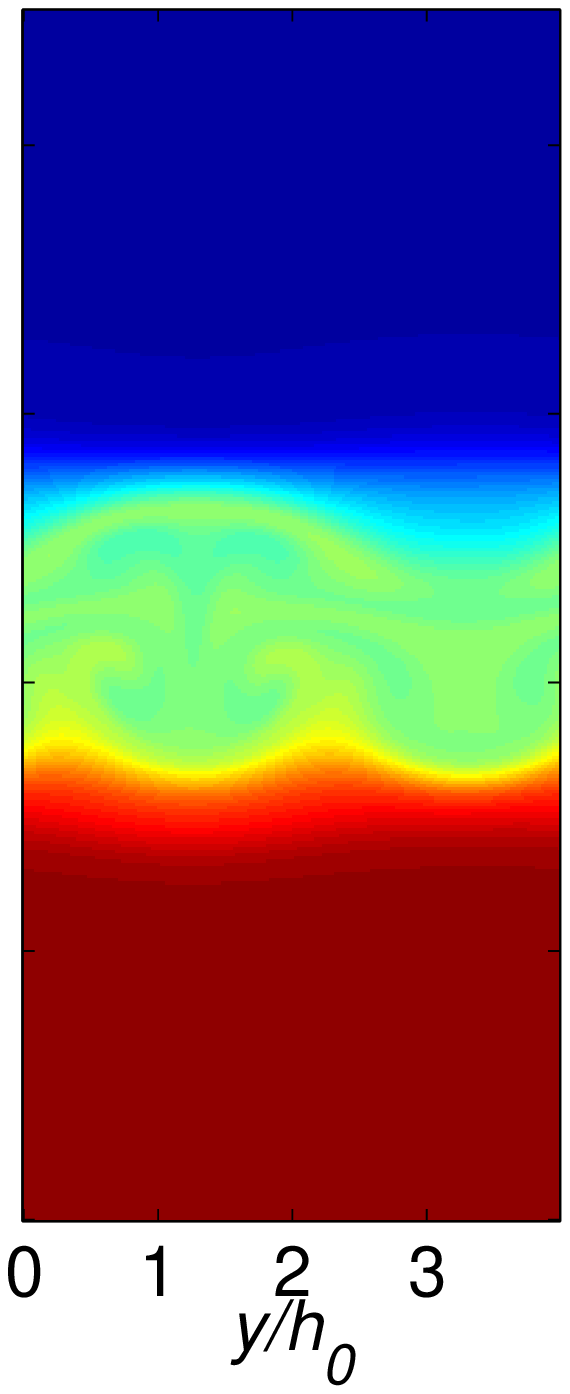}
\end{minipage}
\centering
\hspace{-0.2cm} %
\begin{minipage}[h]{0.18\linewidth} % A minipage that covers half the page
\centering
\includegraphics[height=6.0cm]{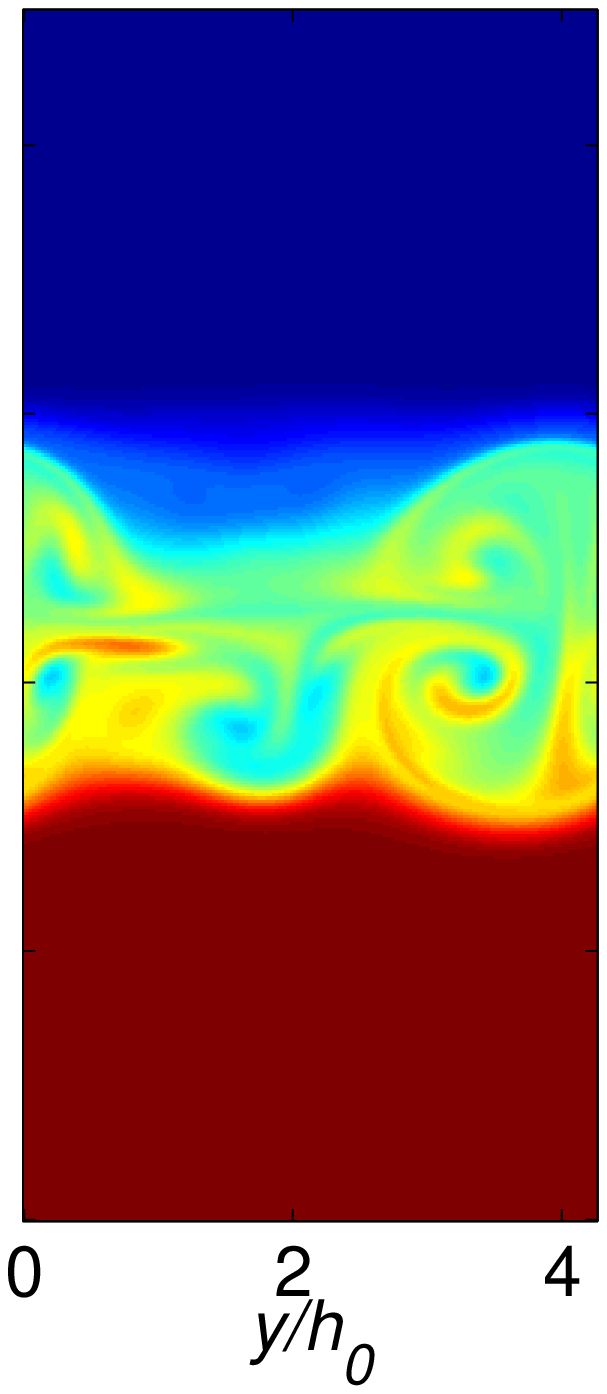}
\end{minipage}
\centering
\hspace{0.05cm} %
\begin{minipage}[h]{0.18\linewidth} % A minipage that covers half the page
\centering
\includegraphics[height=6.0cm]{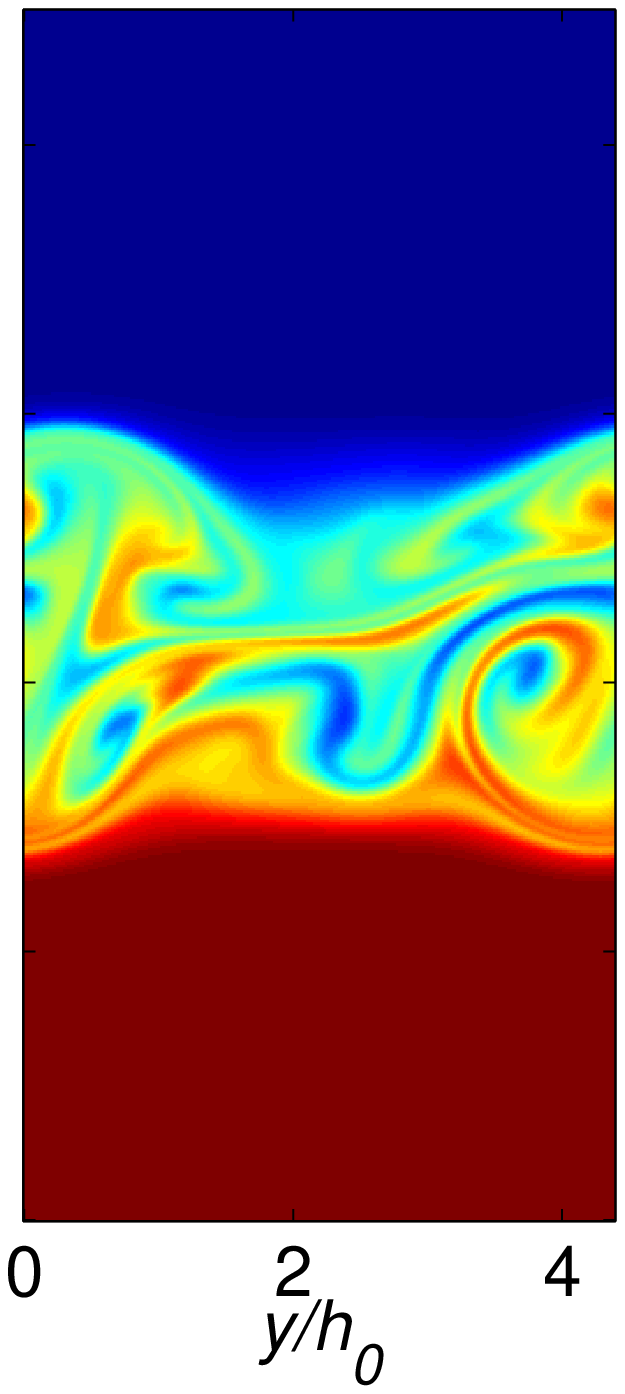}
\end{minipage}
\centering
\hspace{0.05cm} %
\begin{minipage}[h]{0.18\linewidth} % A minipage that covers half the page
\centering
\includegraphics[height=6.0cm]{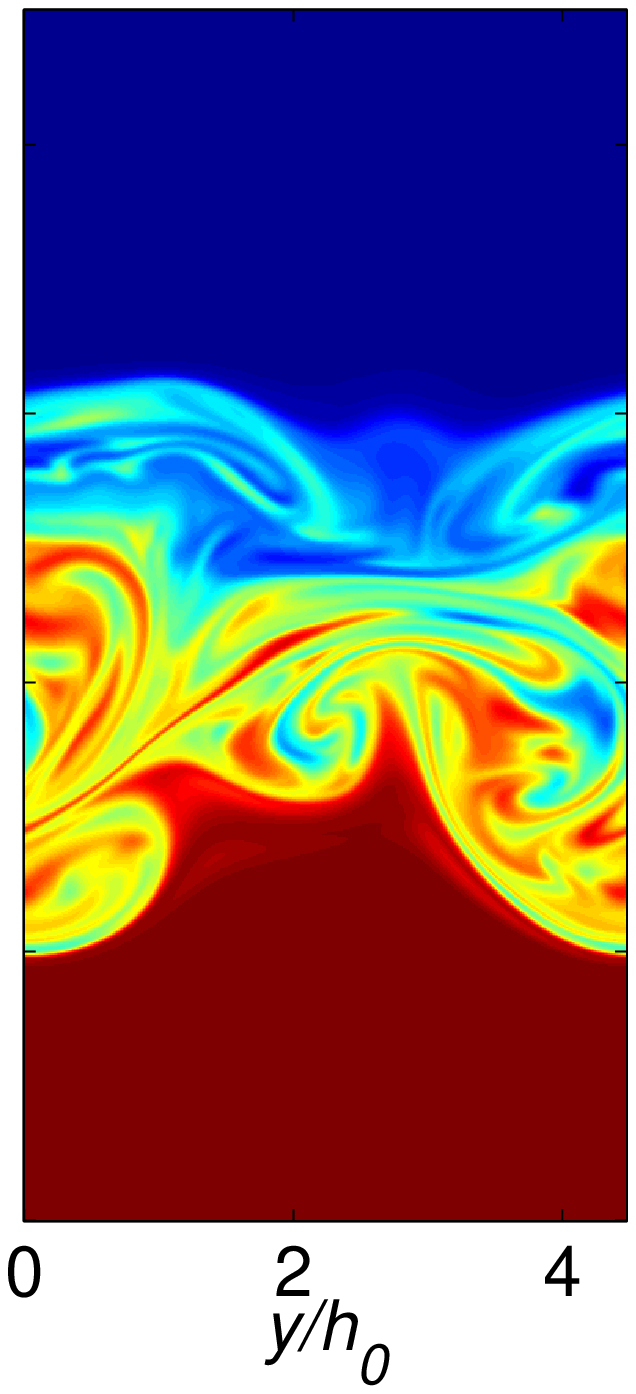}
\end{minipage}

%_______________________________________________________________________________
\hspace{-0.3cm}
\centering
\begin{minipage}[h]{0.18\linewidth} % A minipage that covers half the page
\centering
\includegraphics[height=6.0cm]{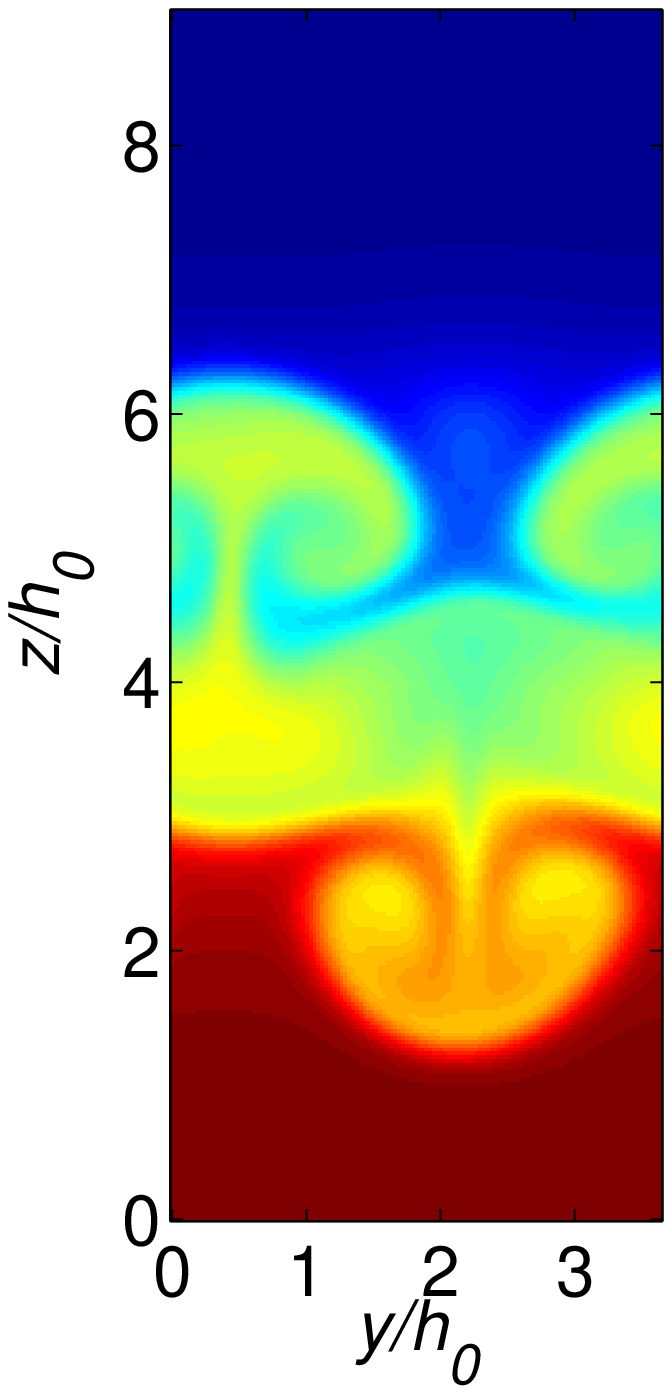}
\end{minipage}
\centering
\hspace{0.4cm} %
\begin{minipage}[h]{0.18\linewidth} % A minipage that covers half the page
\centering
\includegraphics[height=6.0cm]{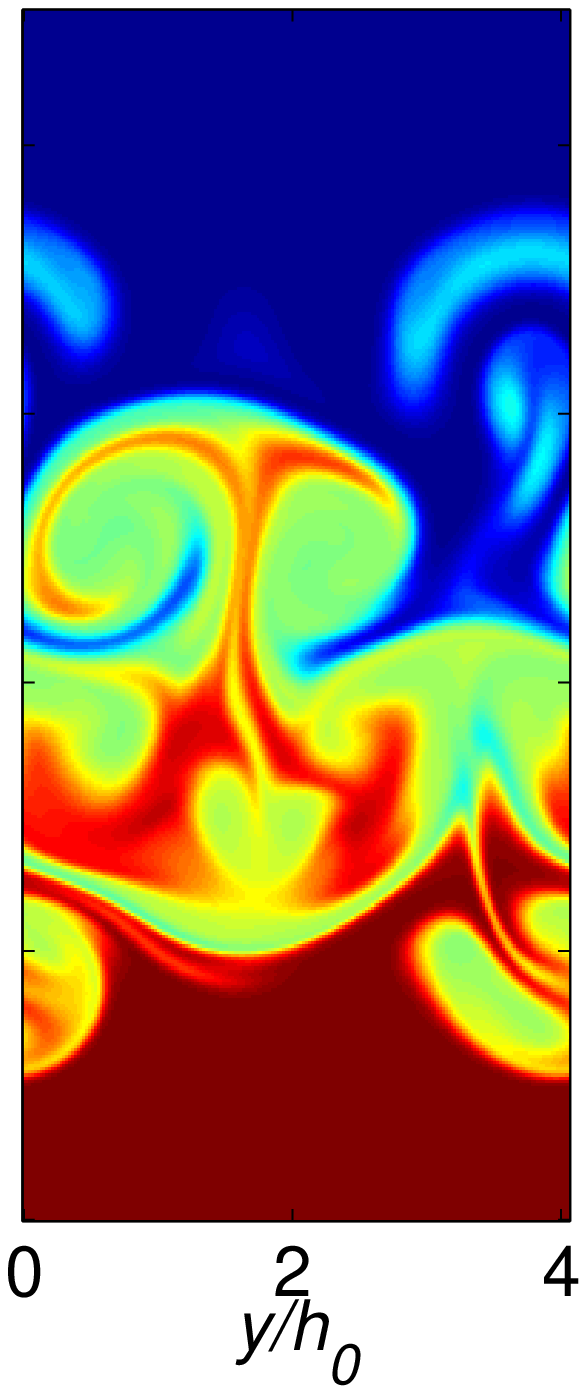}
\end{minipage}
\centering
\hspace{-.2cm} %
\begin{minipage}[h]{0.18\linewidth} % A minipage that covers half the page
\centering
\includegraphics[height=6.0cm]{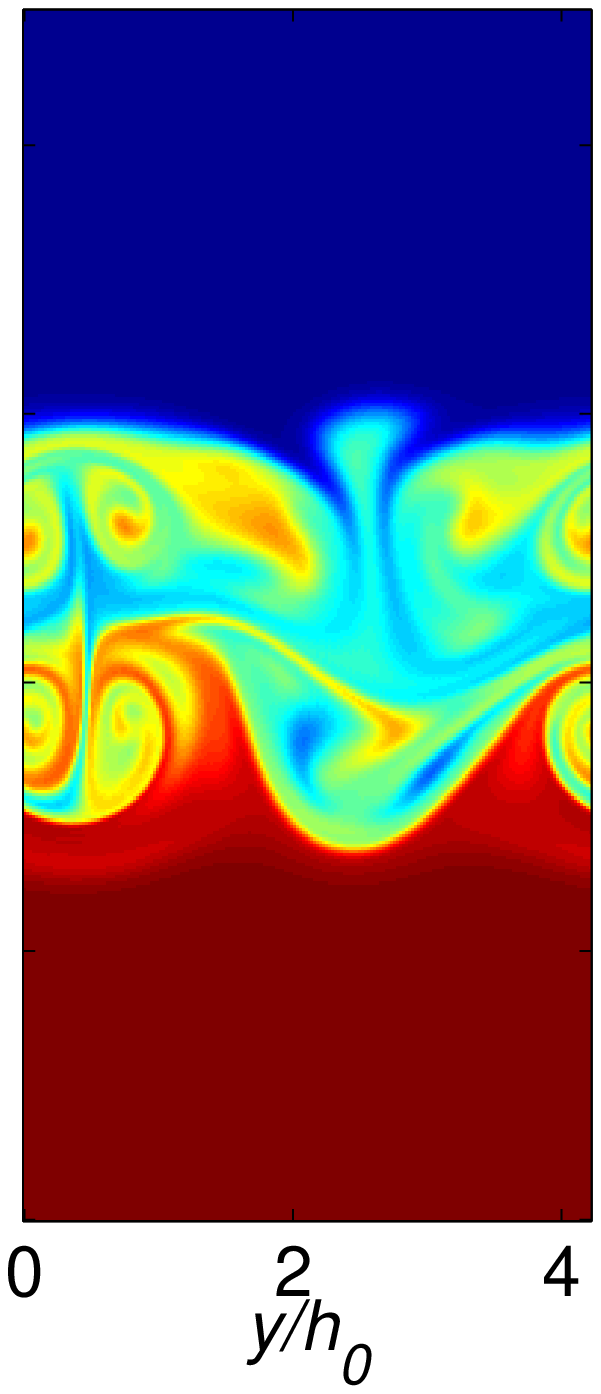}
\end{minipage}
\centering
\hspace{0.1cm} %
\begin{minipage}[h]{0.18\linewidth} % A minipage that covers half the page
\centering
\includegraphics[height=6.0cm]{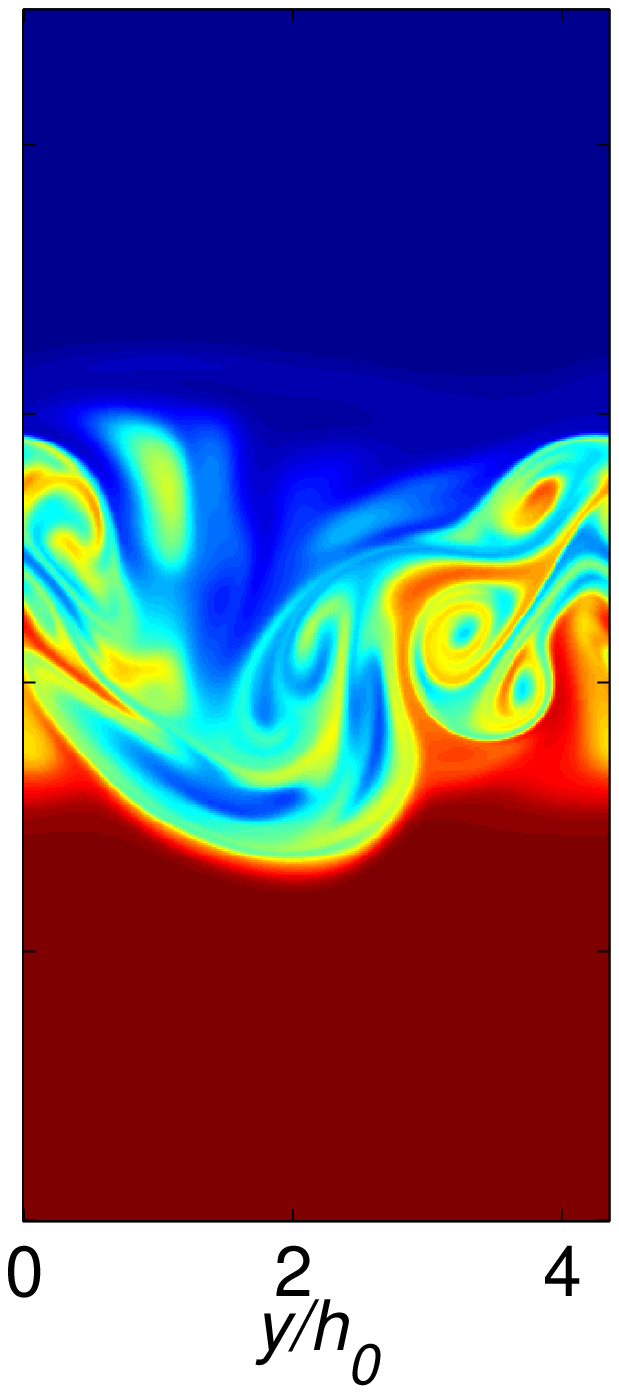}
\end{minipage}
\centering
\hspace{0.1cm} %
\begin{minipage}[h]{0.18\linewidth} % A minipage that covers half the page
\centering
\includegraphics[height=6.0cm]{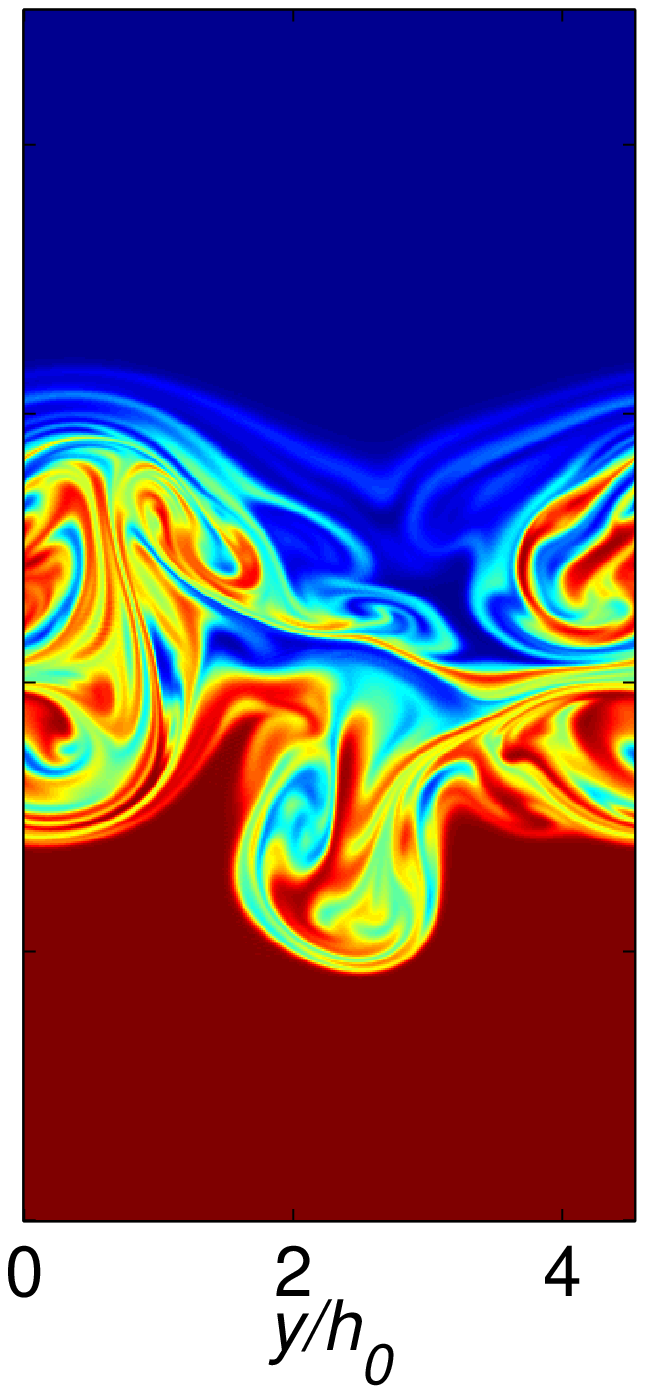}
\end{minipage}
\caption[Snapshots of the span-wise density structure of the billow]{Snapshots of the span-wise density structure of the billow at $x=L_{x}/2$ at $t^{*}_{3}$ for $Re_{0}$ = 300 (top row) and $Pr$ = 1, 9, 16, 25 and 64, from left to right, and $Re_{0}$ = 400 (bottom row) and $Pr$ = 1, 9, 16, 25, and 64 from left to right.}
\label{snapshot_span}
\end{figure}

The structure of the two-dimensional flow in the pre-turbulent stage determines the nature of secondary three-dimensional instabilities and the transition to turbulence, as discussed by \cite{klaassenetal1985a,klaassenetal1985b,klaassenetal1991,caulfieldetal2000,peltieretal2003} . Therefore, the significantly different billow structures in the pre-turbulent stage for different $Pr$ indicates different transition mechanisms to turbulence. Specifically, high density gradients at high $Pr$ make the flow more susceptible to small-scale instabilities. This effect is seen in the span-wise structure of the billow at $t^{*}_{3}$, as shown in figure \ref{snapshot_span}. At $Re_{0}$ = 300 and $Pr$ = 1, three-dimensional motions are mainly suppressed by viscosity and therefore the billow merely diffuses out without becoming turbulent. This case is below the mixing transition. However, as $Pr$ increases for $Re_{0}$ = 300, the three-dimensional motions become stronger and more structured. So as suggested by the two-dimensional simulations in figure \ref{snapshot_stream}, the more structured density field of the pre-turbulent billows at higher $Pr$ enhances three-dimensionality and interfacial area. This effect, attributed to higher baroclinically generated vorticity at sharper density gradients at higher $Pr$, is strong enough to overcome viscous effects and sustain three-dimensional structures through the life-cycle of the billow, even at  the low Reynolds number of $Re_{0}$ = 300. Therefore, the mixing transition occurs at lower critical Reynolds numbers at higher $Pr$.

At $Re_{0}$ = 400, the Reynolds number is high enough for the flow to develop three-dimensional instabilities even at $Pr$ = 1. The flow is within the transitional range according to \cite{breidenthal1981} and \cite{koochesfahanietal1986}.  By $t^{*}_{3}$, the counter-rotating streamwise vortices \cite[][]{linetal1984} have rolled up the fluid into a mushroom-like structure and increased the vertical extent of the billow at all $Pr$. As $Pr$ increases, the density gradients become sharper and smaller scale structures develop in the flow. However, this does not enhance the vertical extent of the flow as much as it did at $Re_{0}$ = 300. This is because the less diffuse density structure at higher $Pr$ is more strongly stratified, and therefore more difficult to displace vertically. Hence, when three-dimensionality is strong, the increase in $Pr$ results in lower buoyancy fluxes and therefore less entrainment, although it locally enhances the intensity of three-dimensional motions. 

\section{Mixing properties}
We quantify mixing, $M$, as the irreversible increase in background potential energy, $P_{b}$, where $P_{b}$ is the minimum potential energy obtained when fluid particles are rearranged adiabatically to a stable state, $\rho(z^{*})$ \cite[][]{wintersetal1995}. In our notation $P_{b}$ is normalized by the initial kinetic energy of the flow, i.e.
\begin{equation}
P_{b}=\frac{g\langle \rho z_{*} \rangle_{V}}{\rho_0 (\triangle U)^{2} }, \; \; \; \; \;  M=P_{b}-P_{b}(0)-\phi_{i}t^{*},
\end{equation}

\noindent , with $\langle  \rangle_{V}$ being the volume average. This quantification of mixing excludes diffusion of the initial background stratification at rate $\phi_{i}$. The rate of mixing is therefore $\phi_{M}=dM/dt^{*}$.  The increase  in the total potential energy, normalized by the initial kinetic energy of the flow, $P$, is a measure of stirring, $T$, i.e.
\begin{equation}
P=\frac{g\langle \rho z\rangle_{V}}{\rho_0 (\triangle U)^{2} }, \; \; \; \; \;  T=P-P(0)-\phi_{i}t^{*}.
\end{equation}

\noindent  The amount of stirring shows how much energy has been extracted from the background shear to vertically (and reversibly) displace the heavier fluid. The difference between $M$ and $T$ is the potential energy available for mixing.  The rate of increase of potential energy is measured by the dimensionless turbulent buoyancy flux, $\Phi^{'}_{b}=(\delta_{0}g/\Delta U^{3}\rho_{0}) \langle w^{'}\rho^{'}\rangle_{V}$. The turbulent buoyancy flux shows how much energy is exchanged reversibly between the background shear and the potential energy.

When studying mixing, it is also useful to examine the dimensionless rate of dissipation of turbulent kinetic energy, $\varepsilon^{'}=(\nu \delta_{0}/ \Delta U^{3}) \langle (\partial u^{'}_{i}/\partial x_{j})^{2} \rangle _{V}$, using tensor notation. While $\phi_{M}$ shows the rate at which the energy extracted from the mean flow converts to mixing, $\varepsilon^{'}$ indicates the rate of dissipation of the extracted energy from the mean flow. The ratio of these two rates, $\phi_{M}/\varepsilon^{'}$, is a commonly used measure of instantaneous efficiency of mixing \cite[e.g.][]{caulfieldetal2000}.

\subsection{Stirring versus mixing}
\begin{figure}
\centering
\includegraphics[width=16.7cm]{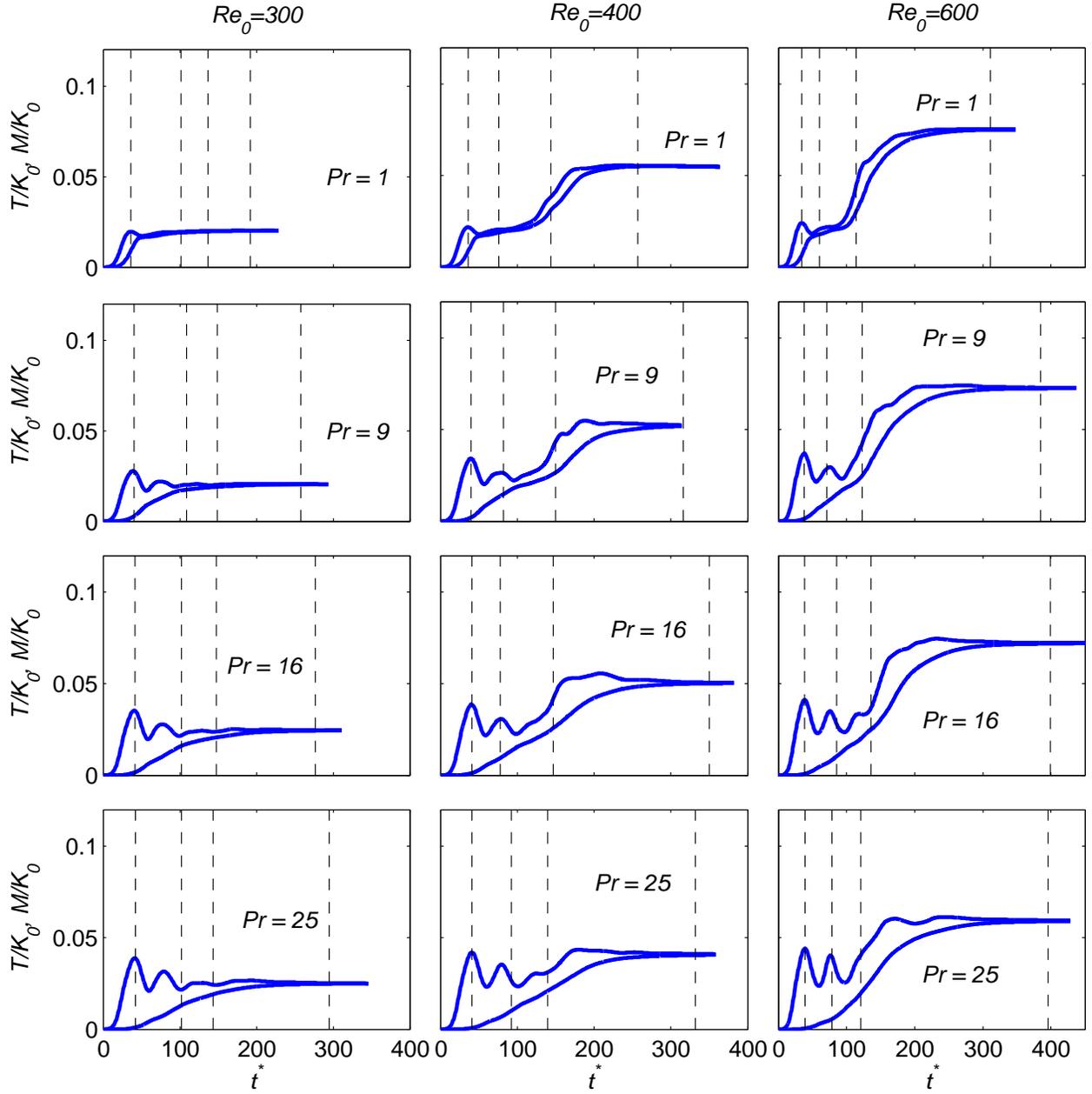}
\caption[Time variation of $T$ and $M$ from three-dimensional simulations]{Time variation of stirring, $T$, (upper solid line) and mixing, $M$, (lower solid line) from three-dimensional simulations. The vertical dashed lines indicate the transition times between the phases.}
\label{fig:T_2_3D_re_300}
\end{figure}

Time variation of stirring, $T$, and mixing $M$, from three-dimensional simulations, for different $Re_{0}$ and $Pr$ is presented in figure \ref{fig:T_2_3D_re_300}. Through the two-dimensional  phases, the buoyancy fluxes act to first increase the stirring, and then generate oscillations in the stirring as the elliptical vortex of the billow oscillates \cite[]{guhaetal2013}. At all $Re_{0}$, the oscillations are more pronounced at higher $Pr$, indicating a more concentrated core vorticity of the billow. The stirring generated during third and fourth phases is suppressed by the increase in $Pr$. As seen in section 3, this is because turbulent buoyancy fluxes are weaker at higher $Pr$, and therefore less unmixed fluid is entrained inside the turbulent region. Moreover, as $Pr$ increases a higher portion of $T$ is exchanged back with the kinetic energy in the turbulence decay phase, without converting into mixing. This is because of the slow rate of mixing at high $Pr$. Mixing occurs with a longer delay after stirring for higher $Pr$. Therefore, the difference between stirring and mixing, the available potential energy, is higher for higher $Pr$. For all $Re_{0}>300$ billows,  a significant portion of mixing occurs in the turbulence decay phase, while the three-dimensional structure of the flow is collapsing. 

\subsection{Rate of mixing and dissipation of kinetic energy}

\begin{figure}
\centering
\includegraphics[width=16.7cm]{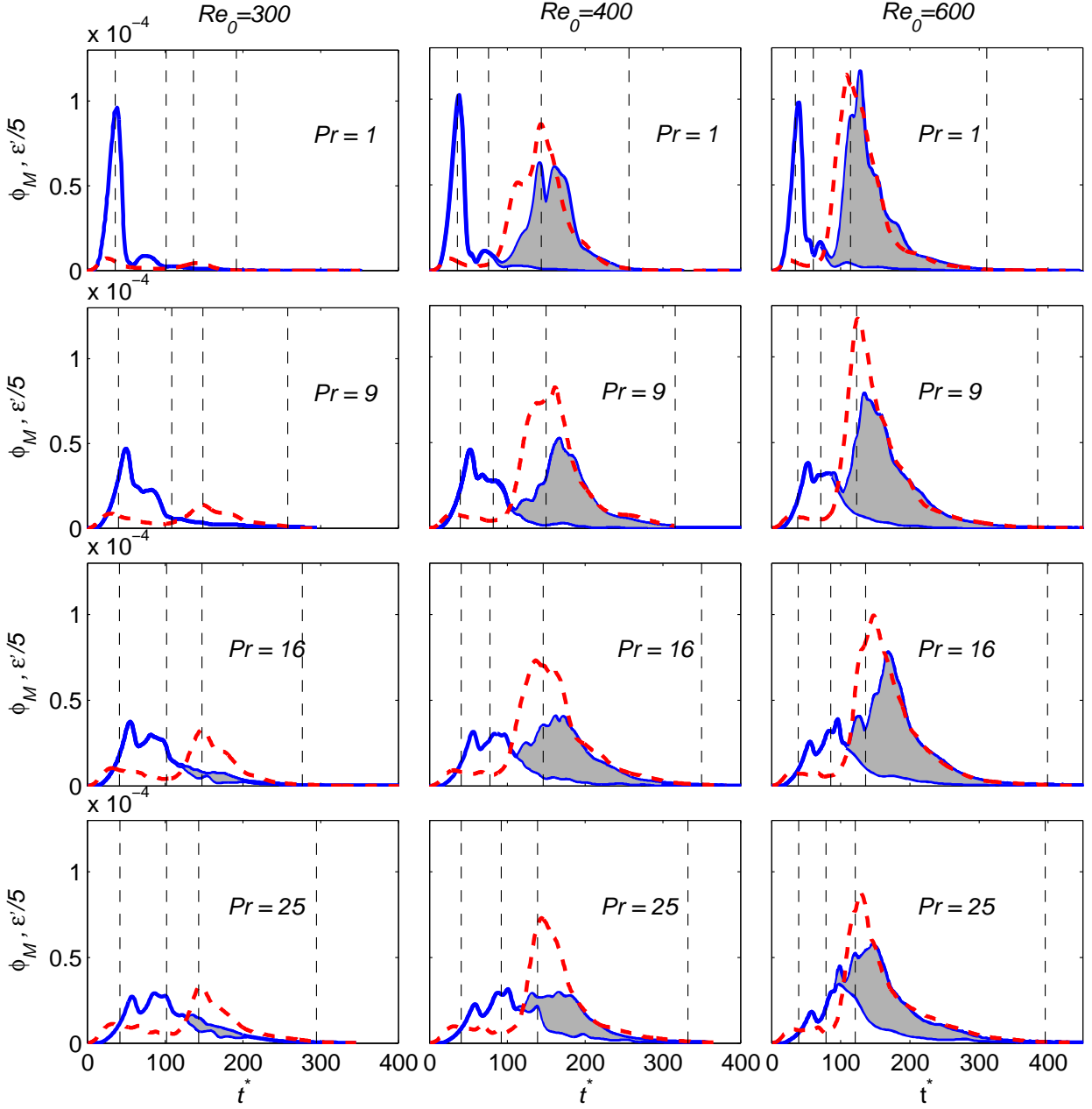}
\caption[Time variation of $\phi_{M}$ and $\varepsilon$ from two- and three-dimensional simulations]{Time variation of the rate of mixing, $\phi_{M}$, from two-dimensional simulations (lower solid line) and three-dimensional simulations (upper solid  line)  and rate of viscous dissipation of kinetic energy (dashed red line) at $Pr$ = 1, 9, 16, and 25 and $Re_{0}$ = 300, 400, and 600. The filled areas show the three-dimensional mixing. The vertical dashed lines indicate the transition times between the phases. For $Re_{0}=300$, at $Pr$ = 1 and 9 the two- and three-dimensional simulations are indistinguishable.}
\label{fig:M_2_3D_re_300}
\end{figure}

Time variation of  $\phi_{M}$ and $\varepsilon^{'}$ through different stages, from two and three-dimensional simulations at $Re_{0}$ = 300, 400, and 600 for different $Pr$, is shown in figure \ref{fig:M_2_3D_re_300}. For $Re_{0}$ = 300, at $Pr$ = 1 and 9, mixing mainly occurs during the first and second phases of the life-cycle, with high rates of mixing. At $Pr$ = 16, and 25, however, mixing occurs at lower rates, and over the whole life-cycle of the billow.  As $Pr$ increases, small-scale and three-dimensional motions are enhanced for this Reynolds number. Therefore, higher shear production of turbulent kinetic energy provides a means for higher extraction of energy from the background flow. This enhances $\varepsilon^{'}$ and three-dimensional $\phi_{M}$ at $Pr$ = 16, and 25. However, the mixing induced by three-dimensional motions is small compared to the total amount of mixing. This is because the enhanced three-dimensional motions are not sufficiently strong for the slow mixing process at high $Pr$ to generate high amounts of mixing. 

At $Re_{0}$ = 400 the flow has sufficient energy to overcome viscous forces and impose strong three-dimensional motions, and therefore the rate of three-dimensional mixing and viscous dissipation of kinetic energy dramatically increases. At $Re_{0}$ = 600, $\phi_{M}$ and $\varepsilon^{'}$ increase more, as the flow is still in the transitional regime. Similar to $Re_{0}$ = 300, the increase in $Pr$ at these two higher Reynolds numbers also prolongs the life-cycle of the billow, and delays the mixing of fluid entrained by the two-dimensional roll-up until third and fourth phases. However, mixing generated by the three-dimensional motions significantly diminishes as $Pr$ rises.  As explained in section 3, the less diffuse density field at higher $Pr$ suppresses buoyancy fluxes and therefore the entrainment in three-dimensional phases. More importantly, while the billow lives long enough for the two-dimensionally rolled-up interface to completely mix, the duration of turbulence is not long enough for high $Pr$ turbulent mixing. Therefore, the three-dimensional structures collapse before the fluid across the enhanced interfacial area is fully mixed. This observation is in agreement with lower rate of turbulent mixing at higher Prandtl numbers observed in turbulent flows by \cite{turner1968,meryfieldetal1998,gargettetal2003,jacksonetal2003}. In the end of the turbulence decay phase for all flows at $Re_{0}$ = 400, and 600, $\varepsilon^{'}/5$ closely follows $\Phi_{M}$, indicating an instantaneous mixing efficiency of close to 0.2, the widely quoted value of turbulent mixing efficiency in the ocean \cite[][]{Smythetal2001, smythetal2007,Inoueetal2009}.

\subsection{Overall amount of mixing}

\begin{figure}
\hspace{-0.8cm}
\centering
\includegraphics[height=5.5cm]{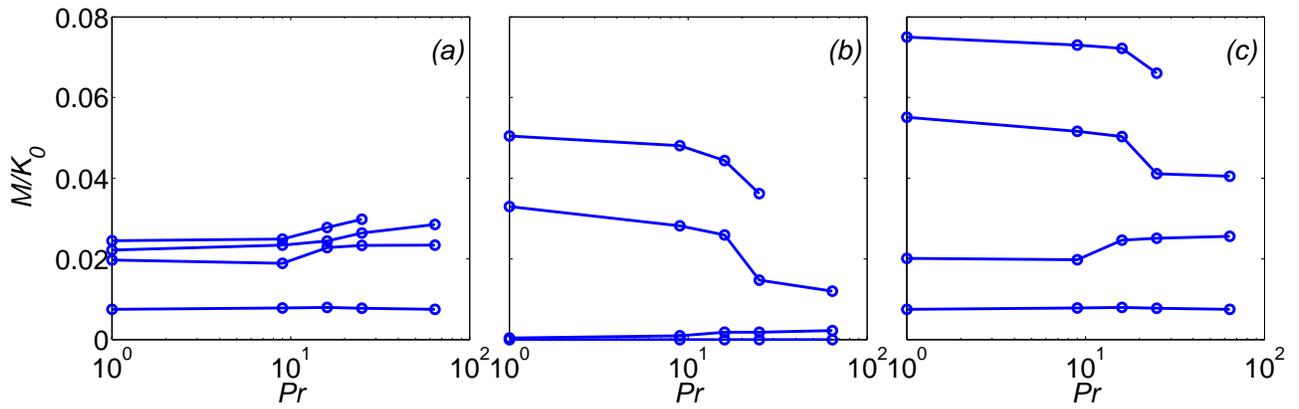}
\centering
\caption[ Variation of the two and three-dimensional, and total amount of mixing with $Pr$]{ Variation of (a) the two-dimensional, (b) the three-dimensional, and (c) the total amount of mixing  with Prandtl number for $Re_{0}$ = 100, 300, 400, and 600 from bottom to top line.}
\label{final_mixing}
\end{figure}

The $Pr$ dependence of the overall two, three-dimensional, and total mixing induced over the entire lifetime of KH billows for different $Re_{0}$ is presented in figure \ref{final_mixing}. The two-dimensional mixing accounts for a significant portion of the total mixing at each of these Reynolds numbers, indicating the significance of the primary vortex in mixing. At $Re_{0}$ = 100, the total amount of mixing does not depend on $Pr$ as viscous effects are dominant. Just before the mixing transition, i.e. $Re_{0}=300$, the total amount of mixing increases with increasing $Pr$,  both due to higher entrainment in the two-dimensional roll-up process, and entrainment by three-dimensional motions. At $Re_{0}$ = 300, the total mixing mainly consists of two-dimensional mixing. For $Re_{0}=$ 400 and 600, while the two-dimensional amount of mixing slightly increases with $Pr$, the three-dimensional and total amount of mixing decreases as $Pr$ increases. The two-dimensional mixing has a weak dependence on $Pr$ as the entrained fluid by the two-dimensional roll-up gets completely mixed inside the billow region. However, as also pointed out by \cite{staquet2000}, $Pr$ has strong effects on the entrainment and mixing of three-dimensional phases of the life-cycle of billows. The three-dimensional mixing exhibits significant amounts for $Re_{0}>300$, above the mixing transition.

\section{Buoyancy and shear Reynolds numbers}
\begin{figure}
\centering
\includegraphics[height=8.3cm]{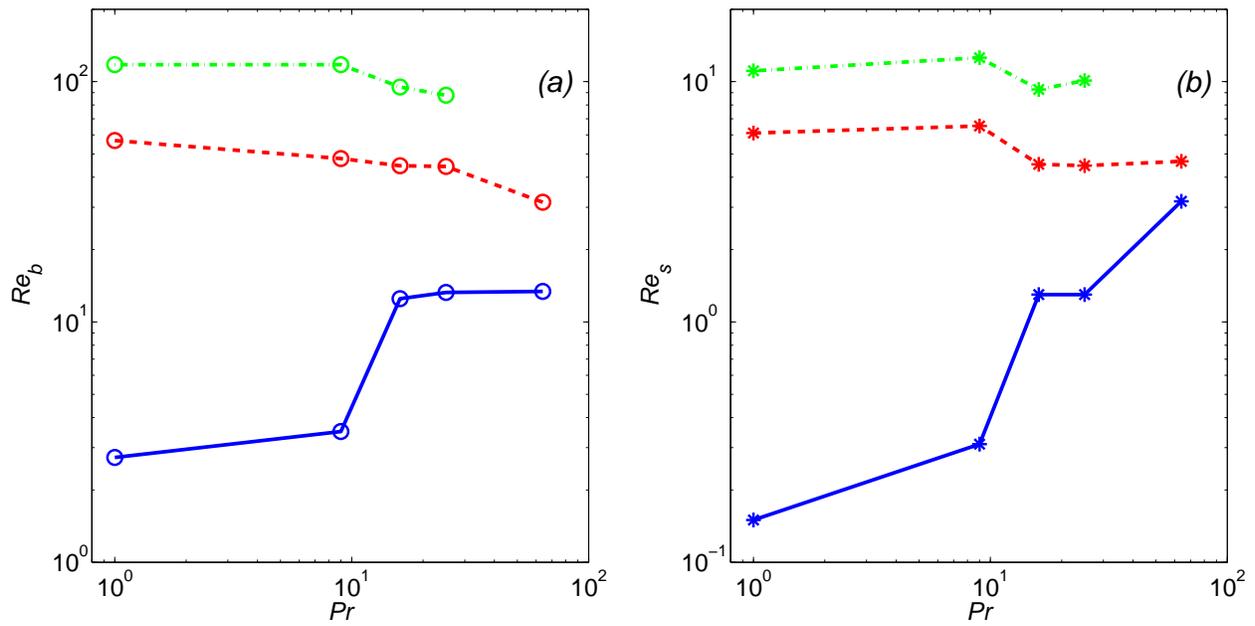}
\centering

\caption[ Variation of buoyancy and shear Reynolds number with $Pr$]{ Variation of  (a) buoyancy Reynolds number ($-o-$), and (b) shear Reynolds number ($-*-$) with Prandtl number for $Re_{0}$ = 300 (solid line) and $Re_{0}$ = 400 (dashed line), and $Re_{0}$ = 600 (dashed dotted line) at maximum intensity of turbulence . }
\label{re_b_s}
\end{figure}

To quantify the effect of stratification and shearing on turbulence and mixing, it is useful to study buoyancy and shear Reynolds numbers, defined as \cite[]{smythetal2000}:
\begin{equation}
Re_{b}= \frac{{\varepsilon}^{'}\Delta U^{3}}{\nu \delta_{0} N^{2}}, \; \; \; \; \;  \textnormal{ and }  \; \; \; \; \;
Re_{s}=\frac{{\varepsilon}^{'}\Delta U^{3}}{\nu \delta_{0} S^{2}} ,
\label{eq5}
\end{equation}

\noindent where $N=(-g\Delta \rho/h \rho_{0})^{1/2}$ is the buoyancy frequency, and $S=\Delta U/\delta$ is the total shear, with $h$ and $\delta$ being the time-varying density and shear layer thickness.  Figure \ref{re_b_s} shows that at maximum intensity of turbulence both $Re_{b}$ and $Re_{s}$ significantly increase as $Pr$ increases at $Re_{0}$ = 300. This indicates that the ratio of the smallest scale distorted  by buoyancy and shear to viscous dissipation scales increases as the flow becomes more actively turbulent and $\delta$, $h$ and ${\varepsilon^{'}}$ grow. At $Re_{0}$ = 400 and 600, $Re_{b}$ and $Re_{s}$ decrease with $Pr$, indicating that buoyancy and shear act over smaller scales as $Pr$ increases. The buoyancy Reynolds numbers in for $Re_{0}=$ 400 and 600 are greater than 20, the critical value for the start of active turbulence \cite[]{smythetal2000}, and  smaller than 100-300, the critical range for the significance of double diffusion \cite[]{smythetal2005,jacksonetal2003}.  Figure \ref{re_b_s} also suggests that the range of maximum $Re_{b}$ and $Re_{s}$ in our simulations is the same as those in simulations of \cite{smythetal2000}, where they considrered higher initial Reynolds numbers, but stronger stratification. The range of $Re_{b}$ in the measurements of \cite{moum1996} in the ocean thermocline was between $O(10^{2})$ and $O(10^{4}$). Therefore, some of our simulations represent the lower range of buoyancy Reynolds numbers in real oceanic processes. \ 

\begin{figure}
\centering
\includegraphics[height=12.5cm]{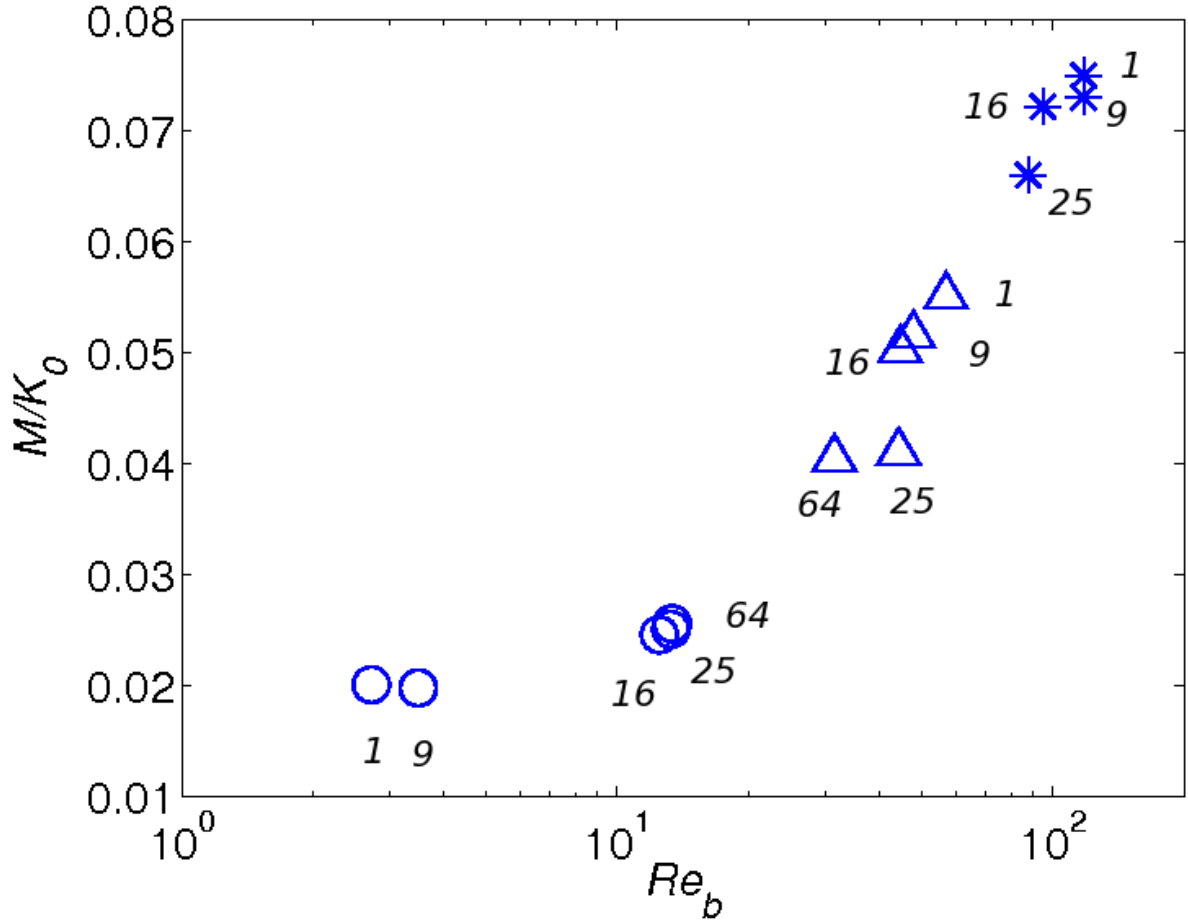}
\centering
\caption[ Variation of the total amount of mixing with $Re_{b}$]{ Variation of the total amount of mixing  with maximum buoyancy Reynolds number. The markers indicate different initial Reynolds numbers with different Prandtl numbers: $Re_{0}$ = 300 (circles), $Re_{0}=400$ (triangles), and $Re_{0}=600$ (stars). The number next to each data point shows the Prandtl number.}
\label{mixing_reb}
\end{figure}

Finally, we present the variation of the overall amount of mixing with maximum $Re_{b}$ in figure \ref{mixing_reb}. The buoyancy Reynolds numbers incorporates both effects of $Re_{0}$ and $Pr$ into one parameter. The amount of mixing increases with $Re_{b}$, with a transition at the rate of increase for $Re_{b} >10$. This is the buoyancy Reynolds number above which our simulations exhibit active three-dimensional motions. The monotonic increase of the amount of mixing with $Re_{b}$ indicates that when turbulent motions are less distorted by buoyancy, more mixing is generated. While the increase in $Re_{0}$ inhibits buoyancy effects, the increase in $Pr$ inhibits or enhances buoyancy effects depending on whether the flow is within, or below the mixing transition.

\section{Conclusions}

Two and three-dimensional DNS were performed to examine the effect of $Pr$ on the evolution and overall amount of mixing in KH billows. The increase in $Pr$ has significant implications for the strength of turbulent three-dimensional motions, buoyancy fluxes, and mixing. These effects are through enhancing the density gradients and interfacial area at higher $Pr$. The $Pr$ dependence of the final amount of mixing shows two distinct behaviours for Reynolds numbers below and within the mixing transition. For $Re_{0}$ = 300, below the mixing transition, three-dimensional motions and interfacial area are enhanced as $Pr$ rises, and as a result the mixing increases. For $Re_{0}>300$, however, the slower rate of diffusion dominated over the effect of enhanced interfacial area at higher $Pr$, and the overall mixing decreases. The two-dimensional entrainment is not significantly dependent on $Pr$. The two-dimensional mixing adjusts its lifetime with the slower mixing at higher $Pr$, and therefore the overall two-dimensional mixing is weakly dependent on $Pr$. However, the three-dimensional mixing occurs over a limited time due to the dissipative nature of turbulence. Therefore, three-dimensional mixing is repressed by slower mixing at high $Pr$. The effects of initial Reynolds number, and Prandtl number on buoyant fluxes and mixing are well captured in the buoyancy Reynolds number, and an increasing trend of amount of mixing with increasing buoyancy Reynolds number is observed, with a transition at critical buoyancy Reynolds number of 10.\\

We are thankful to Dr. Kraig Winters and Dr. Bill Smyth for providing the DNS code, and Westgrid for computational resources.

%%%%%%%%%%%%%%%%%%%%%%%%%%%%%%%%%%%%%%%%%%%%%%%%%%%%%%%%%%%%%%%%%%%%%
% FIGURES
%%%%%%%%%%%%%%%%%%%%%%%%%%%%%%%%%%%%%%%%%%%%%%%%%%%%%%%%%%%%%%%%%%%%%

%__________________________________________
\begin{table}
\centering
  \begin{tabular}{ccccccccccc} %\\ %\hline
	 % 3 dimensional
	Simulation & $\textit{Pr}$  &  $\textit{R}$  &$L_{x}/\delta_{0}$&$L_{y}/\delta_{0}$& $N_{x}\times N_{y}\times N_{z}$ & $\Delta z/L_{B}$\\
	\hline
        $Re_{0}$ = 100\\
        \hline	
	1&1&1&8.3& 4.15&256$\times$128$\times$256 &0.3\\
        2&9&3&9.1&4.55 &256$\times$128$\times$256 &1.0\\
        3&16&4&9.5&4.75 &256$\times$128$\times$256 &1.4\\
        4&25&5&9.8& 4.9&256$\times$128$\times$256&1.8\\
        5&64&8&10.5& 5.25&256$\times$128$\times$256&2.6\\        \hline	
        $Re_{0}$ = 300\\
        \hline	
        6&1&1&7.5& 3.75&256$\times$128$\times$320& 0.4\\
        7&9&3&8.0&4.0 &384$\times$192$\times$384&1.1\\
        8&16&4&8.6& 4.3&384$\times$192$\times$384&1.7\\
        9&25&5&8.8& 4.4&512$\times$256$\times$512& 1.6\\
        10&64&8&9.1&4.55 &512$\times$256$\times$512&2.5\\
        11&700&26.5&9.8& 0&2000$\times$0$\times$2000&-\\
	\hline
        $Re_{0}$ = 400\\
        \hline	
	12&1&1&7.4&3.7 &256$\times$128$\times$320&1.0\\
        13&9&3&8.2& 4.1&384$\times$192$\times$384&1.9\\
        14&16&4&8.5&4.25 & 384$\times$192$\times$384&2.4\\
        15&25&5&8.7&4.35 &640$\times$320$\times$640&1.8\\
        16&64&8&9.1 &4.55 & $768\times$ 384$\times$768&2.4 \\       
         \hline	
	$Re_{0}$ = 600\\
        \hline	
	17&1&1&7.3& 3.65&256$\times$128$\times$320&1.4\\
        18&9&3&8.1&4.05 &480$\times$240$\times$480&2.2\\
        19&16&4&8.4& 4.2&512$\times$256$\times$512&2.6\\
        20&25&5&8.6& 4.3&640$\times$320$\times$640&2.6\\
        \hline	
        	\\     %\hlin
		\end{tabular}
			\caption[Description of the two and three-dimensional simulations]{Description of the two and three-dimensional simulations carried out. In all these simulations $J$ = 0.03, $L_{z} = 9\delta_{0}$. Simulations 6 to 16 have also been performed in two dimensions, with $L_{y} = 0$. The presented mesh sizes are for the density field, the velocity field mesh sizes are half of these. } %Take out comments? Mention duplicates? 
	\label{tab:sim}

\end{table}

%______________________________________________________________

%______________________________________________________________

%________________________________________________________________________________________

%________________________________________________________________________________________

%________________________________________________________________________________________

%________________________________________________________________________________________

%________________________________________________________________________________________

%%%%%%%%%%%%%%%%%%%%%%%%%%%%%%%%%%%%%%%%%%%%%%%%%%%%%%%%%%%%%%%%%%%%%
% REFERENCES
%%%%%%%%%%%%%%%%%%%%%%%%%%%%%%%%%%%%%%%%%%%%%%%%%%%%%%%%%%%%%%%%%%%%%
% Create a bibliography directory and place your .bib file there.
\ifthenelse{\boolean{dc}}
{}
{\clearpage}
\bibliographystyle{ametsoc}
\bibliography{references}
%\bibliography{bib_thesis}

\end{document}